\documentclass[a4paper,11pt]{article}
\pdfoutput=1 

\usepackage{jheppub} 
\usepackage{epstopdf} 
\usepackage[T1]{fontenc} 
\usepackage{bbm}

\usepackage{graphics}
\usepackage{inputenc}
\usepackage{xspace}
\usepackage{amsmath}
\usepackage{amssymb}
\usepackage{url}
\usepackage{rotating}
\usepackage{enumerate}
\usepackage{graphicx}
\usepackage{def} 

\title{\boldmath Color matrix element corrections for parton showers}

\author[a]{Simon Pl\"atzer}
\author[b]{Malin Sjodahl}
\author[b]{Johan Thor\'en}

\affiliation[a]{Particle Physics, Faculty of Physics, University of Vienna}
\affiliation[b]{Department of Astronomy and Theoretical Physics, Lund
  University, S{\"o}lvegatan 14A, 223\,62 Lund, Sweden}

\emailAdd{simon.plaetzer@univie.ac.at}
\emailAdd{malin.sjodahl@thep.lu.se}
\emailAdd{johan.thoren@thep.lu.se}

\abstract{We investigate the effects of keeping the full color structure for
  parton emissions in parton showers for both LEP and LHC.
  This is done within the Herwig 7 dipole shower, and includes gluon emission, gluon
  splitting, initial state branching processes, as well as hadronization.
  The subleading $\Nc$ terms are included as color matrix element corrections to the
  splitting kernels by
  evolving an amplitude-level density operator and correcting the
  radiation pattern for each parton multiplicity, up to a fixed number of full color
  emissions, after which a standard leading color shower takes over.
  Our results are compared to data for a wide range of LEP and LHC
  observables and show that the subleading $\Nc$ corrections tend to be small
  for most observables probing hard, perturbative dynamics, for both LEP and LHC.
  However, for some of these observables they exceed 10\%.
  On soft physics we find signs of significantly larger effects.
}

\begin{document} 
\preprint{LU-TP 18-21, MCNET-18-14, UWTHPH-2018-21}
\maketitle
\flushbottom

\section{Introduction}
\label{sec:introduction}

At present time event generators
\cite{Sjostrand:2007gs,Bahr:2008pv,Bellm:2015jjp,Gleisberg:2003xi}
have developed into indispensable tools for understanding collider
phenomenology. At the same time, the high energy available at the LHC
has significantly opened up the perturbative phase space available for
radiation.  This increases the demand of resummation, performed either
analytically or, as in the case of parton showers, numerically {\it e.g.}
via the Sudakov veto algorithm \cite{Platzer:2011dq,Platzer:2011dr}.

From a QCD perspective, the high number of colored partons
due to the large perturbative phase space, as well as due
to the fact that the initial state partons carry color, calls for a
better understanding of subleading $\Nc$ effects.

This is also in demand to ease the cancellation of infrared
singularities in the matching and merging of parton showers with NLO
calculations \cite{Platzer:2011bc,Platzer:2012bs,Bellm:2017ktr}, which often fully
include subleading $\Nc$ effects. It should also be
stressed that we expect such corrections to account for
subleading $\Nc$, leading-logarithmic problems which can arise in
dipole-type parton shower algorithms due to an ambiguous definition of
an emitter's color charge \cite{Dasgupta:2018nvj}.

To achieve greater accuracy and better understanding of parton shower
uncertainties, it is therefore time to include a better description of
subleading color contributions in parton showers, similar to how
parton showers recently have been improved by matrix element merging
at leading order \cite{Lonnblad:2001iq,Krauss:2002up,Hoche:2006ph,
  Lavesson:2007uu,Hoeche:2009rj,Hamilton:2009ne}, and matching at
next-to-leading order \cite{Dobbs:2001dq,Frixione:2002ik,
  Nason:2004rx,Nagy:2005aa,Frixione:2010ra,Platzer:2011bc}.

The first steps in this direction has already been performed by some
of the authors in the case of an $e^+e^-$-collider
\cite{Platzer:2012np}.  Others have pursued another road, keeping only
a subset of the color suppressed terms
\cite{Nagy:2012bt,Nagy:2015hwa}, and detailed studies have been
carried out towards systematically expanding virtual and
real effects in shower-type evolution algorithms
\cite{Platzer:2013fha,Martinez:2018ffw}.

In the present paper we extend the color matrix element corrections,
first implemented in \cite{Platzer:2012np}, by including initial
state hadrons as well as $g\to \qqbar$-splittings,
subsequent leading color showering and hadronization.  This is done
within the Herwig 7.1 \cite{Bellm:2017bvx} implementation of the
dipole shower algorithm \cite{Platzer:2009jq}, giving us a
full-fledged general purpose event generator which can be used for
studying color matrix element corrections to any process occurring at
the LHC and other colliders, in practice up to a limited number of
colored partons, restricted by the fast growing complexity in color
space, however still reaching down to relatively soft emissions.

Our method is based on dipole factorization
\cite{Catani:1996jh,Catani:1996vz}, which we outline in
\secref{sec:dipole}. The complication brought about by the color
matrix element corrections is discussed in \secref{sec:cmec}, and
implementation details, involving evolution of the color structure
treatment, the weighted Sudakov algorithm and the density operator,
are discussed in \secref{sec:color}, \ref{sec:sudakov} and
\ref{sec:density_operator} respectively. Results for various
processes, including initial and final state radiation, standard QCD
observables, heavy quark production and $Z$ plus jets are discussed in
\secref{sec:results}, and concluding remarks are made in
\secref{sec:conclusion}.

\section{Essence of dipole factorization and dipole shower evolution}
\label{sec:dipole}

This paper is based on dipole factorization, stating that whenever the next
gluon to be emitted from an $n$-parton configuration becomes either soft or
collinear to one of the existing partons, the squared amplitude for the
$n+1$-parton case can be approximated with
\begin{multline}
  \label{eq:dipolefactorization}
  |{\cal M}_{n+1}(...,p_i,...,p_j,...,p_k,...)|^2 \approx\\
  \sum_{k\ne i,j} \frac{1}{2 p_i\cdot p_j}
  \langle {\cal M}_n(...,p_{\tilde{ij}},...,p_{\tilde{k}},...) |
          {\mathbf V}_{ij,k}(p_i,p_j,p_k)| {\cal M}_n(...,p_{\tilde{ij}},...,p_{\tilde{k}},...)\rangle \ ,
\end{multline}
in terms of the old amplitude $|{\cal M}_{n}\rangle$. In the above, an emitter
$\tilde{ij}\to i,j$ whereas a spectator $k\to\tilde{k}$ absorbs the
longitudinal recoil, such that all partons, before and after emission, stay
on-shell. For final state radiation we use the standard Sudakov decomposition,
\begin{eqnarray}
p_i & = & z p_{\tilde{ij}} +\frac{p_\perp^2}{z s_{ijk}}p_{\tilde{k}} + k_\perp \\
p_j & = & (1-z) p_{\tilde{ij}} +\frac{p_\perp^2}{(1-z) s_{ijk}}p_{\tilde{k}} - k_\perp \\
p_k & = & \left(1-\frac{p_\perp^2}{z(1-z)s_{ijk}}\right)p_{\tilde{k}} \;,
\end{eqnarray}
with $p_{\tilde{ij}}^2=p_{\tilde{k}}^2=0$, a spacelike transverse momentum
$k_\perp$ with $k_\perp^2=-p_\perp^2$ and $k_\perp\cdot p_{\tilde{ij}}=
k_\perp\cdot p_{\tilde{k}}= 0$.  The cases of initial state emitter or
spectator are discussed in \cite{Platzer:2009jq}.

Note that the sum in \eqref{eq:dipolefactorization} only runs over
$\tilde{ij}\ne \tilde{k}$. This is possible since the collinear singularity
corresponding to the square of the diagram where parton $\tilde{ij}$ is the
emitter, has been rewritten as a sum of interferences between that diagram and
every other diagram using color conservation
\begin{equation}
  \label{eq:cc}
        {\mathbf T}_{\tilde{ij}}^2 = - \sum_{k\ne \tilde{ij}} {\mathbf T}_{\tilde{ij}}\cdot {\mathbf T}_k \ .
\end{equation}
The splitting kernels are given in terms of the standard dipole splitting
kernels as
\begin{eqnarray}
  {\mathbf V}_{ij,k}(p_i,p_j,p_k) = 
  -8\pi\alpha_s V_{ij,k}(p_i,p_j,p_k) 
  \frac{{\mathbf T}_{\tilde{ij}}\cdot {\mathbf T}_k}{{\mathbf T}_{\tilde{ij}}^2}\ ,
  \label{eq:Vij}
\end{eqnarray}
where ${\mathbf T}_{\tilde{ij}}\cdot {\mathbf T}_k$ describes the color space
effect of exchanging a gluon between parton $\tilde{ij}$ and parton
$\tilde{k}$, and thus contains an implicit sum over gluon indices.  In the
massless case, the final-final dipole splitting kernels are given by
\begin{eqnarray}
V_{qg,k}(p_i,p_j,p_k) &=& \cf \left( \frac{2 (1-z)}{(1-z)^2+p_\perp^2/s_{ijk}} - (1+z) \right)\nonumber\\
V_{gg,k}(p_i,p_j,p_k) &=& 2C_A\left( \frac{1-z}{(1-z)^2+p_\perp^2/s_{ijk}} + 
\frac{z}{z^2+p_\perp^2/s_{ijk}} - 2 + z(1-z) \right)\ .
\label{eq:VqgVgg}
\end{eqnarray}
Note, however, that dipole factorization is valid also for massive particles
in the quasi-collinear limit, and our implementation is general, using the
massive dipole splitting kernels and kinematics available in Herwig, as
detailed in \cite{ttbarPaper}.
For the dipole configurations involving initial state partons,
we use the corresponding expressions as given in
\cite{Catani:1996vz,Catani:2002hc}. In \eqref{eq:VqgVgg} the factors of
$\cf=\TR(\Nc^2-1)/\Nc$ and $\ca=2 \TR\Nc$, where $\TR$ is defined by $\tr(t^at^b)=\TR\delta^{ab}$, 
explains the inclusion of ${\mathbf T}_{\tilde{ij}}^2$ in
\eqref{eq:Vij}; in order to use the standard definition of
\eqref{eq:VqgVgg}, ${\mathbf T}_{\tilde{ij}}^2$ is introduced in the
denominator of \eqref{eq:Vij}.

In the large $\Nc$ limit the color correlator becomes
\begin{equation}
  -\frac{{\mathbf T}_{\tilde{ij}}\cdot {\mathbf T}_k}{{\mathbf T}_{\tilde{ij}}^2}\to
  \frac{1}{1+\delta_{\tilde{ij}}}\delta(\tilde{ij}, k\text{ color connected})\ ,
  \label{eq:InfCol}
\end{equation}
where $\delta_{\tilde{ij}}\equiv 1$ if $\tilde{ij}$ is a gluon, and zero
otherwise. In this way, the factor $\cf$ for gluon emission off a
quark is reproduced in the large $\Nc$ limit by coherent emission
from the quark and its color-connected partner, and the factor
$\ca$ is reproduced for gluon splitting by the sum of the coherent
emissions from the gluon and its two color-connected partners, hence
the factor $1/2$ for gluons in \eqref{eq:InfCol}.

We remark that the leading $\Nc$ version of the above describes how to get a
leading $\Nc$ correct emission pattern.  It does not address the issue of
assigning a leading $\Nc$ color flow from which subsequent emission can be
performed.  The default strategy is to use a color flow where the radiated
gluon is inserted between the emitter and the spectator. However, with a gluon
splitting kernel which is symmetric in $z\leftrightarrow1-z$, the question of
which gluon is the radiator and which is radiated arises. In the limit
where $z$ is small, it is rather the parton with momentum fraction $z$ which
is soft and thus should be seen as radiated and therefore be inserted between
the emitter (with the large momentum fraction $1-z$) and the spectator.  For
this reason, to mimic the swap of color, we swap momenta of the emitter and
the emitted parton with probability $1-z$, such that the soft gluon always
tends to be inserted between the harder parton and the spectator in the color
structure, guaranteeing that we get the correct soft limit. We remark, however
that the probability $1-z$ is a choice, and that and any function, having the
same limits when $z\to1$ and $z\to0$, would have been a valid choice.
Alternatively, the splitting kernels could have been redefined to only contain
the $1-z$ singularity (see \cite{Gustafson:1987rq,Bellm:2018wwz} for comparison).  

As we want to describe LHC collisions, the color matrix element corrections
have to be applied also to initial state hadrons, meaning that we have to deal
with initial-initial emissions as well as initial-final and final-initial
emissions. The initial state emission cases are treated in a standard backward
evolution scheme, meaning that if the emitter is an initial state parton, the
backward evolution is done by folding in the parton distribution functions
(PDFs), using appropriate splitting kernels, and colors are updated as if the
resulting (low energy) parton was emitted.  To be more precise, denoting the
emitter participating in the hard process by $i$, the radiated parton with
$j$, and the (initial) parton going into the PDF by $\tilde{ij}$, the used
splitting kernel is $P_{\tilde{ij}\to ij}$, and the emission probability is
evolved using PDF ratios \cite{Sjostrand:1985xi}, as described in section 6.5 in \cite{Bahr:2008pv}.
The color structure of the full color shower, on the other hand, is, as discussed in
\secref{sec:density_operator}, treated as if $i \to \tilde{ij}, j$.

In the shower algorithm presented here, we
also extend our analysis from \cite{Platzer:2012np} by including $g\to\qqbar$
splitting. The description of how this splitting fits into the dipole
formalism is given in \secref{sec:density_operator}.

\section{Color matrix element corrections}
\label{sec:cmec}

We like to stress that this paper deals with color matrix element
corrections to parton showers. We thus correct each emission with
the full color correlations, keeping all soft and collinear
contributions to the emission, i.e., we use the right "antenna pattern".
In this sense, we do more than the standard leading $\Nc$ showers ---
where only the leading $\Nc$-terms, and the color suppressed term
in $\cf=\TR(\Nc^2-1)/\Nc$, are kept --- but less than a full matrix
element correction. 

We thus start from some color amplitude, $|{\cal M}_n\rangle$,
for $\Np$ colored partons,
decomposed in any arbitrary color basis, with basis vectors
$|\alpha_n\rangle$

\begin{equation}
|{\cal M}_n\rangle = \sum_{\alpha=1}^{d_n} c_{n,\alpha} |\alpha_n\rangle
\quad \leftrightarrow \quad {\cal M}_n = (c_{n,1},...,c_{n,d_n})^T \ .
\end{equation}
Squaring the amplitude in order to calculate a cross section gives
\begin{equation}
\label{eq:m2}
|{\cal M}_n|^2 = {\cal M}_n^\dagger S_n {\cal M}_n =
\Tr \left( S_n\times {\cal M}_n{\cal M}_n^\dagger \right) 
\end{equation}
with $S_n$ being the scalar product matrix, 
$(S_n)_{\alpha\beta}=\langle\alpha_n|\beta_n\rangle$,
for color basis vectors $|\alpha_{\Np} \rangle$ and $|\beta_{\Np} \rangle$.
Upon emission from the $\tilde{ij},\tilde{k}$-pair, the relevant color
factor changes to
\begin{equation}
\label{eq:ccm2}
\langle {\cal M}_n|{\mathbf T}_{\tilde{ij}}\cdot {\mathbf T}_{\tilde{k}}|{\cal M}_n\rangle = 
\Tr \left( S_{n+1}\times T_{\tilde{k},n} {\cal M}_n{\cal M}_n^\dagger T_{\tilde{ij},n}^\dagger \right)
\end{equation}
in terms of matrix representations,
${T}_{\tilde{ij}},{T}_{\tilde{k}}\in {\mathbb C}^{d_{n+1},d_n}$, 
of
${\mathbf T}_{\tilde{ij}},{\mathbf T}_{\tilde{k}}$. 

While this improves the radiation pattern to fixed order, we should remark 
that it is not the full story of color suppressed terms in the context of
parton showers. To fully include all subleading $\Nc$ terms in the soft
and collinear limits, virtual color rearranging terms associated with
the same singularity structure should also be kept. To accomplish this,
a full resummation of virtual exchanges is needed. Unfortunately, within
the current event generator structure these contributions cannot be
included, and we postpone their inclusion for future work.
Instead we present a fully functional subleading $\Nc$ dipole shower,
building on the algorithm presented in \cite{Platzer:2012np}, but including
$g \to \qqbar$ splitting, hadronization, full mass dependence and initial
state hadrons, meaning final-final, initial-final, final-initial and initial-initial dipoles.

The rewriting of the collinear singularities in terms of dipole splitting
kernels eqs.~(\ref{eq:Vij}-\ref{eq:VqgVgg})\xspace, using color conservation,
\eqref{eq:cc}, means that the color flow will be replaced by every other color
flow, except the flow associated with the color structure of the collinear
singularity.  In order to be able to continue the full $\Nc$ matrix element
corrected parton shower with a leading $\Nc$ parton shower (as well as with
subsequent hadronization), we do, however, keep the color structure associated
with emission from the emitter, as in the standard event record, and we use
that for the subsequent leading $\Nc$ shower and hadronization. Not having one
color structure to start from when it comes to hadronization would imply
considering a new approach to a hadronization model, which is much beyond the
scope of the current paper. On the other hand, it is still interesting to add
hadronization in a standard way to enable comparison to data.

\section{Color structure treatment}
\label{sec:color}

So far, the treatment of color structure in this paper has been completely
basis independent. Any complete spanning set of relevant color structures,
such as trace bases \cite{Paton:1969je, Berends:1987cv, Mangano:1987xk,
  Mangano:1988kk, Kosower:1988kh, Nagy:2007ty, Sjodahl:2009wx, Alwall:2011uj,
  Sjodahl:2014opa, Platzer:2012np}, multiplet bases
\cite{Kyrieleis:2005dt,Dokshitzer:2005ig,Sjodahl:2008fz,Beneke:2009rj,
  Keppeler:2012ih, Du:2015apa, Sjodahl:2015qoa}, or color flow bases
\cite{'tHooft:1973jz,Kanaki:2000ms,Maltoni:2002mq} would do, as long as the matrices $T_{\tilde{ij},n}$ for
gluon emission, the matrices $t_{\tilde{ij},n}$ describing gluon splitting and the
scalar product matrices $S_n$ could be calculated.  This basis freedom is
maintained within the Matchbox module of Herwig, which can be used to
interface to any implementation of color structure, such as CVolver
\cite{Platzer:2013fha} or ColorFull \cite{Sjodahl:2014opa}.

For our simulations we use trace bases and the ColorFull implementation
\cite{Sjodahl:2014opa}.  In the trace bases, the color structure is expressed
in terms of open and closed quark-lines, of the form
\begin{eqnarray}
  \label{eq:color_example}
  \parbox{5.5 cm}{\includegraphics[width=5.5cm]{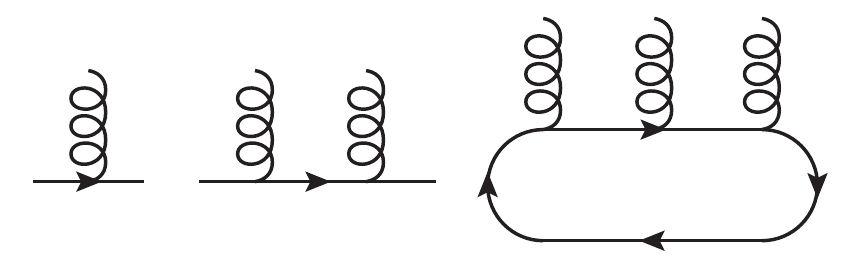}}
\end{eqnarray}
for all possible quark and gluon permutations and all possible number of
traces \cite{Sjodahl:2009wx,Sjodahl:2014opa}.  These bases have several
advantages.  The color structure can trivially be translated into the leading
$\Nc$ color flow, which is needed for subsequent leading $\Nc$ emission and
standard hadronization.  The processes of gluon emission, gluon splitting, and
gluon exchange can all be very easily described, giving respectively at most
two, two and four new basis vectors
\cite{Mangano:1988kk,Nagy:2007ty,Sjodahl:2009wx}.  For example, considering
gluon emission off an open quark-line we have the two trace basis vectors
\begin{equation}
  \label{eq:gluon_emission}
  \parbox{4cm}{\includegraphics[width=4cm]{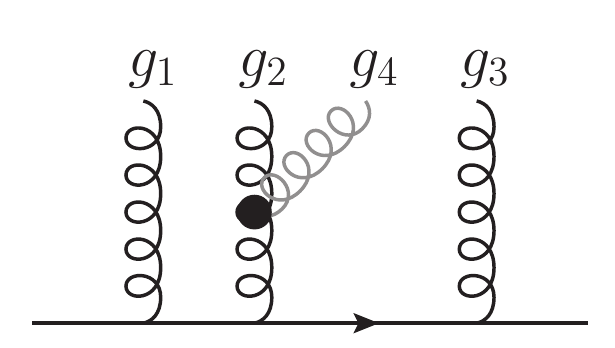}}
  =
  \parbox{4cm}{\includegraphics[width=4cm]{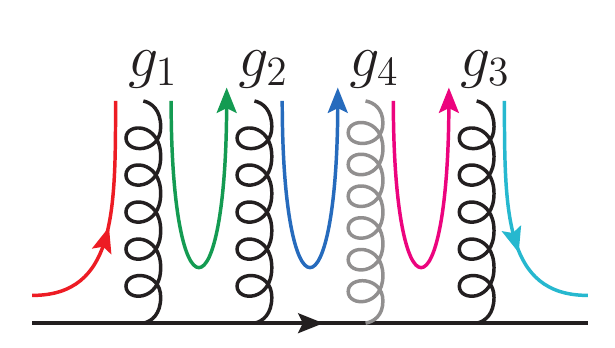}}
  -
  \parbox{4cm}{\includegraphics[width=4cm]{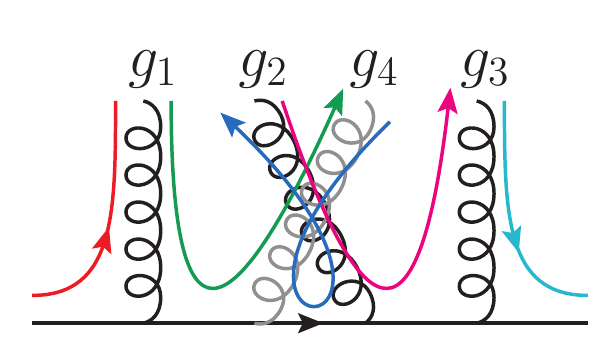}}\ ,
\end{equation}
where we have also illustrated the different color flows in the
$\Nc\to \infty$ limit.\footnote{The relative sign on the right hand
  side is a matter of convention, and must be matched with the sign
  of the kinematics structure. Here we apply ColorFull's
  convention of introducing a minus sign when the emitted gluon
  is inserted before the emitter on the quark-line.}
If we instead consider gluon splitting, we have 

\begin{equation}
  \label{eq:gluon_split}
  \parbox{4cm}{\includegraphics[width=4cm]{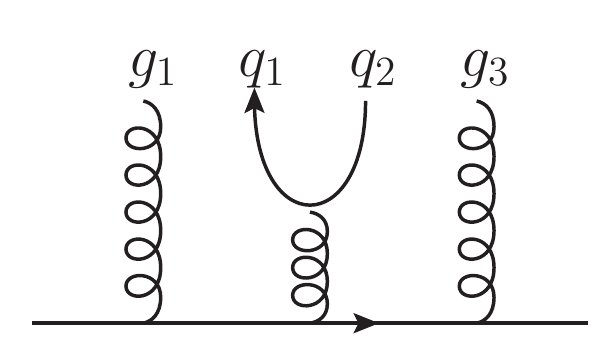}}
  =\TR \left[
  \parbox{4cm}{\includegraphics[width=4cm]{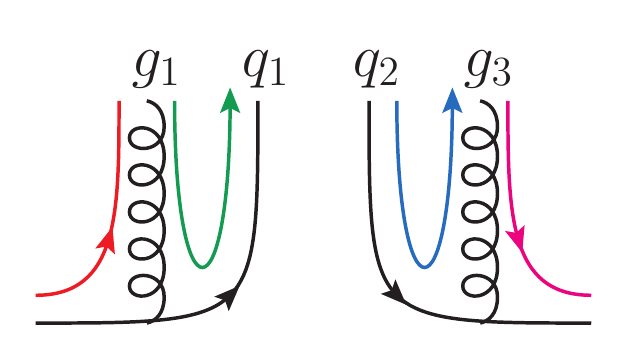}}
  -
  \frac{1}{\Nc}
  \parbox{4cm}{\includegraphics[width=4cm]{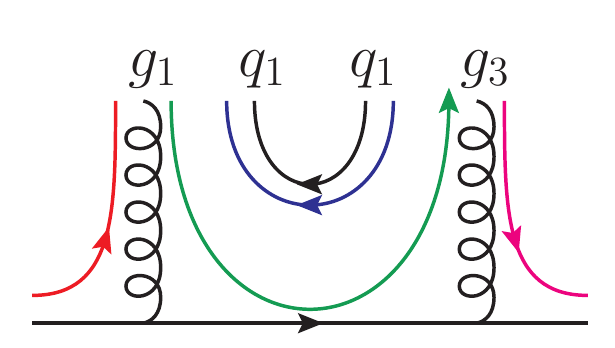}}\right]\ .
\end{equation}
One of the disadvantages of trace basis lies in their
overcompleteness, meaning that hence --- strictly speaking
--- they are actually not bases, but rather spanning sets.

For a small set of partons, up to approximately
five $\qqbar$-pairs and gluons, the overcompleteness is
fairly moderate, but for more partons it becomes
significant \cite{Keppeler:2012ih,Sjodahl:2015qoa}.
The number of basis vectors scales as a factorial in the
trace basis case, with roughly $(\Ng+\Nqqbar)!/e$
basis vectors \cite{Keppeler:2012ih}, whereas the number of basis vectors for
finite $\Nc$ scale only as an exponential. On the other hand,
basis vectors which are not needed for a given process can often
easily be identified and crossed out for trace bases,
in particular, at tree-level, the number of required basis
vectors scales approximately as $(\Ng+\Nqqbar-1)!$.
For example, in \eqref{eq:color_example}, only color structures
where all gluons are attached to the open quark-lines contribute.

The overcompleteness, and the fact that the scalar product matrices
are dense, i.e., the scalar product between most pairs of
basis vectors does not vanish, is the Achilles' heel of the
trace bases.
Instead of vanishing for two different basis vectors, the
scalar product is suppressed by one or more powers
of $1/\Nc$. Thus, when $\Nc \to \infty$, the bases are orthogonal,
and the color structure can be replaced by color flows, as
in \eqref{eq:gluon_emission} and \eqref{eq:gluon_split}.
Since we keep the full color structure,
it is the calculation of scalar products that limits
the number of subleading $\Nc$ emissions that we can keep.
At about $\Ng+\Nqqbar=6$, the color structure calculations
start to take up significant time, and going beyond
$\Ng+\Nqqbar=7$ is very time consuming within our current
setup.

This unfavorable scaling behavior could be circumvented by using orthogonal
multiplet bases \cite{Keppeler:2012ih, Sjodahl:2015qoa}, in which case the
calculation of the radiation matrices $T_{\tilde{ij},n}$ would be more time
consuming, however not to the same degree \cite{Du:2015apa}.  Alternatively
color structure could be sampled over.  This is a road which we have attempted
to pursue. Within our current framework, using the weighted Sudakov veto algorithm
from \cite{Bellm:2016voq} (modified as described in \secref{sec:sudakov}), 
it is, however, impractical to go beyond 3 subleading $\Nc$ emissions at
LEP, or two subleading $\Nc$ emissions at LHC, due to the very poor statistical
convergence from large weight fluctuations. Therefore, within our current implementation,
sampling has proven disadvantageous.

\section{The weighted Sudakov algorithm}
\label{sec:sudakov}
The shower described in this paper treats up to $\Nmax$ emissions with the
full color correlations. This corrects the emissions to appear first in the
$p_\perp$-ordered evolution, down to smaller scales, and then the leading
color shower handles the subsequent lower $p_\perp$ emissions. The radiation
pattern we aim at describing in the full color shower is
\begin{equation}
{\rm d}P_{ij,k}(p_\perp^2,z;p_{\tilde{ij}},p_{\tilde{k}}) = 
\frac{\alpha_s}{2\pi}\frac{{\rm d}p_\perp^2}{p_\perp^2} {\rm d}z
{\cal J}(p_\perp^2,z;p_{\tilde{ij}},p_{\tilde{k}}) V_{ij,k}(p_\perp^2,z;p_{\tilde{ij}},p_{\tilde{k}})
\times \frac{-1}{{\mathbf T}_{\tilde{ij}}^2} \frac{\langle
  {\cal M}_n|{\mathbf T}_{\tilde{ij}}\cdot {\mathbf T}_{\tilde{k}} |{\cal M}_n\rangle
}{|{\cal M}_n|^2},
\label{eq:dP}
\end{equation}
where the factor after the multiplication sign is the color matrix element
correction, which we will denote by $\omega_{\tilde{ij}\,\tilde{k}}^n$. For
the cases when the color matrix element correction is negative, the weighted
Sudakov veto algorithm from \cite{Bellm:2016voq} is used, with one
modification: In the competition version of the algorithm from \cite{Bellm:2016voq}, the weight for an
emission gets a contribution from each veto and accept step, for all trial
emissions of all competing pairs. The algorithm will, however, produce the same distribution if the
total weight only receives contributions from the accept step of the winning
emission and veto steps at scales larger than the winning scale. Discarding
all weight contributions below some scale (in this case the scale of the
winning emission) makes the algorithm generate another distribution below that
scale, but the radiation pattern of the losing trial emissions below the
winning scale cannot affect the final distribution, since these emissions
are discarded anyway. 
However, the convergence of
the algorithm gets worse, due to large weight fluctuations.
We therefore choose to discard the weights below the winning scale.
This modification significantly improves the convergence.

Two choices in the algorithm from \cite{Bellm:2016voq}
are: the acceptance probability (denoted $\epsilon$ in
\cite{Bellm:2016voq}) and the overestimate proposal distribution (denoted
$R$). These free choices were used to improve the convergence of the algorithm,
the details of the choices can be found in \appref{app:sudakov}. 

In total, to reproduce the radiation pattern in \eqref{eq:dP}, the shower
steps given below are repeated until $\Nmax$ emissions have been corrected or
no emission is found above the cut-off scale $\mu$.
\begin{enumerate}
\item The starting scale $Q_\perp$ is given by the hard scale. For the
  first emission, this is taken to be the $Z$ mass for LEP and the average
  transverse momentum of hard jets in the final state for LHC. 
  For subsequent emissions, $Q_\perp$ is given by the scale of the previous
  emission.
\item All processes for all pairs of partons compete with each other and a
  winning hardest scale is chosen. For each dipole, $\tilde{ij},\tilde{k}$,
  candidate emissions, $\tilde{ij},\tilde{k}\to i,j,k$ at scales
  $p_{\perp,\tilde{ij},k}$, are chosen according to the Sudakov form factor
\begin{equation}
-\ln \Delta_{ij,k}(p_{\perp,ij,k}^2|Q_\perp^2) =
\frac{\alpha_s}{2\pi}\int_{p_{\perp,ij,k}^2}^{Q_\perp^2} \frac{{\rm d}q_\perp^2}{q_\perp^2}
\int_{z_-(q_\perp^2)}^{z_+(q_\perp^2)}{\rm d}z\ 
{\cal P}_{ij,k}(q_\perp^2,z;p_{\tilde{ij}},p_{\tilde{k}}) \ ,
\end{equation}
where ${\cal P}_{ij,k}$, in accordance with \eqref{eq:dP}, is
\begin{equation}
  \label{eq:Pijk}
{\cal P}_{ij,k}(p_\perp^2,z;p_{\tilde{ij}},p_{\tilde{k}})
={\cal J}(p_\perp^2,z;p_{\tilde{ij}},p_{\tilde{k}}) V_{ij,k}(p_\perp^2,z;p_{\tilde{ij}},p_{\tilde{k}})
\times
\frac{-1}{{\mathbf T}_{\tilde{ij}}^2} \frac{\langle
  {\cal M}_n|{\mathbf T}_{\tilde{ij}}\cdot {\mathbf T}_k |{\cal M}_n\rangle
}{|{\cal M}_n|^2}
\end{equation}
and $z_\pm(p_\perp^2)$ follow from the phase space boundaries at
fixed transverse momentum. If the color matrix element correction is positive,
the standard Sudakov veto algorithm is used (resulting in that the trial
emission always contributes a factor $1$ to the event weight) and if it is
negative, the modified weighted veto algorithm is used (where the weight is,
in general, multiplied by the weight in \eqref{eq:veto}). The winning emission defines the details of
the kinematics and the recoil is absorbed by the spectator $\tilde{k}$ of the
winning dipole, such that all partons are on-shell after the emission.
\item If no scale above the cut-off $\mu$ was found, the shower terminates.
\item If this is the emission $\Nmax$, the leading $\Nc$ shower will continue
  showering the event, otherwise the density operator is updated as will be
  described in \secref{sec:density_operator}.
\end{enumerate}
If $\Nmax$ emissions have been corrected, the leading $\Nc$ shower continues
with the color structure given by the large $\Nc$ flow associated with
emissions from the selected emitters, as discussed in \secref{sec:cmec}. The
leading $\Nc$ shower then continues until reaching the cut-off scale
$\mu$. Finally, the event may, or may not, be hadronized.  If the
hadronization is performed it starts from the leading $\Nc$ color flow.

\section{Evolution of the density operator}
\label{sec:density_operator}

The parton shower starts from the hard matrix element $|{\cal M}_n \rangle$.
However, after emission, the resulting ``dipole'' color structure from
\eqref{eq:ccm2} cannot, within our framework, be cast into the form of some
new amplitude $|{\cal M}_{n+1} \rangle$. Instead we see from \eqref{eq:ccm2}
that the relevant $n+1$-parton quantity, corresponding to $M_n\equiv{\cal
  M}_n{\cal M}_n^\dagger$ for emission from $\tilde{ij},\tilde{k}$ is $\sim
T_{\tilde{k},n} {\cal M}_n{\cal M}_n^\dagger T_{\tilde{ij},n}^\dagger$.  For
gluon emission (final or initial), keeping all contributions to the emission
probability, and using the dipole factorization \eqref{eq:dipolefactorization}
with the splitting kernels \eqref{eq:Vij}, we see that if we define
\begin{equation}
\label{eq:nextamplitude}
M_{n+1} =
-\sum_{i\ne j}\sum_{k\ne i,j} \frac{4\pi\alpha_s}{p_i\cdot p_j} \frac{V_{ij,k}(p_i,p_j,p_k)}{{\mathbf T}_{\tilde{ij}}^2}
\ T_{\tilde{k},n}M_n T_{\tilde{ij},n}^\dagger \ ,
\end{equation}
the matrix element square for $\Np+1$ particles can be written analogously to
\eqref{eq:m2} as
\begin{equation}
\label{eq:mn+12}
|{\cal M}_{n+1}|^2 = 
\Tr \left( S_{n+1}\times M_{n+1}\right). 
\end{equation}
Thus, when a phase space point has been selected for gluon emission, $M_n$
could be updated according to \eqref{eq:nextamplitude}.  Note, however, that
the overall normalization of $M_{n+1}$ is irrelevant, since when used in the
$\Np+2$-version of \eqref{eq:Pijk} to calculate the emission of $\Np+2$
partons, $M_{n+1}$ enters in both the numerator and denominator. Thus we could
ignore any constant factor.  In fact, for technical reasons, we only keep the
eikonal parts, $\sim p_i \cdot p_k/(p_i\cdot p_j\, p_k\cdot p_j) $, of \eqref{eq:nextamplitude}. Clearly the dropped hard collinear pieces
should not alter the subsequent emission of soft wide-angle radiation.

For the case of $g\to\qqbar$, there is no interference between various
possible emitters, and the amplitude is symmetric in all final state gluons,
meaning that $M_n$ can be updated using only one term
\begin{equation}
  \label{eq:nextamplitudegqqbar}
  M_{n+1} =t_{\tilde{ij},n}M_n t_{\tilde{ij},n}^\dagger \ ,
\end{equation}
where $t_{\tilde{ij},n}$ represents the color space map corresponding
to $t^g_{\qqbar}$, i.e., the matrix where element $\alpha\beta$ is the
transition from basis vector $\beta$ in the initial (smaller) basis,
to basis vector $\alpha$ in the final (larger) basis, where the
difference between the color structures is that the gluon $\tilde{ij}$
has been contracted and replaced by the $\qqbar$-pair $i,j$, giving
one or two new basis vectors. The possible recoil partners used to set
  up the gluon splitting into quarks are picked using \eqref{eq:dP} with
  the $g\to \qqbar$ splitting kernel, but with color matrix element corrections {\it as for gluon emission}.
  While this choice is ad-hoc in this case, it has
  the advantage of nicely fitting into the dipole picture.
In the collinear limit, where the
splitting becomes relevant, we can use color conservation \eqref{eq:cc},
to observe that the sum over the kernels using different spectators is indeed
collapsing to the expected collinear splitting function.

Note that the same update of $M_{n+1}$, \eqref{eq:nextamplitudegqqbar}, and
the same recoil strategy, is applied irrespectively of if the splitting gluon
is final or initial. The only difference is thus, as for the gluon emission
case, the standard convolution with the PDFs for initial parton splitting
rates. If we have an initial state quark (antiquark) which evolves backwards
into a gluon going into the parton distribution function and an antiquark
(quark) which is radiated, the shower has no interference with other diagrams,
and the density matrix can be updated according to
\begin{equation}
  \label{eq:nextamplitude2}
  M_{n+1} =T_{\tilde{ij},n}M_n T_{\tilde{ij},n}^\dagger \ ,
\end{equation}
without an irrelevant, overall factor.  In the standard dipole large $\Nc$
shower, the momentum recoil is absorbed by the color-connected partner,
implying that the emission can be accounted for within the dipole shower
formalism.

A comment on the update of the density matrix is in order here. In this paper
we keep the full color structure, of all possible emitter-spectator pairs
which could have contributed to an emission. An alternative strategy is to
pairwise sample over emitters and spectators, corresponding to keeping only
one term in the double sum in \eqref{eq:nextamplitude}.  While this samples
from all color structures, we like to remark that it actually corresponds to a
slightly different approximation. In our current implementation, after defining the
kinematics of the new emission from the winning pair, {\it all} pairs contribute to
the color structure of the next emission, and the weight of their color
structure, is given by \eqref{eq:nextamplitude}, implicitly multiplying the
same Sudakov factor, the Sudakov factor of the winning pair. In a sampling
procedure, the terms in the \eqref{eq:nextamplitude} would end up in different
events, and each term would be associated with the Sudakov factor {\it of the pair
emitting} in that winning phase space point. The difference in a sufficiently
large Monte Carlo sample, thus lies within the Sudakov factor, coming with
different scales for different pairs.
We note
however, that the two approaches should agree in the soft limit, and we have
also checked that the numerical difference is small.

\section{Results}
\label{sec:results}

\subsection{Outline of the simulation}

In this section we outline the simulation, and consider
various shower distributions
with the aim of understanding and validating the shower evolution.

We use the Herwig 7.1 dipole shower, with settings according to the 7.1.3
release \cite{Bellm:2017bvx}, with the modified weighted Sudakov veto
algorithm outlined in \secref{sec:sudakov}, and with color matrix element
corrections as described in \secref{sec:cmec}, starting from lowest order
$2\to2$ processes.  The LHC generation $p_\perp$-cut is by default put to 30
GeV, and the default jet analysis veto is $p_\perp=50$ GeV.  The energy is 13
TeV for LHC and 91.2 GeV for LEP unless stated otherwise. If jet clustering
is required, our default choice is the anti-$k_T$ algorithm, as provided by
the fastjet package \cite{Cacciari:2008gp}, with $R=0.4$ at LHC.
  At LEP, no generation cuts are applied, and
  both at LEP and LHC we use the original Rivet \cite{Buckley:2010ar} analysis published along with
  the data in comparison to data.

To ensure statistical convergence, we start with investigating the
weight distribution.
\begin{figure}
  \centering
  \includegraphics[width=0.45\textwidth]{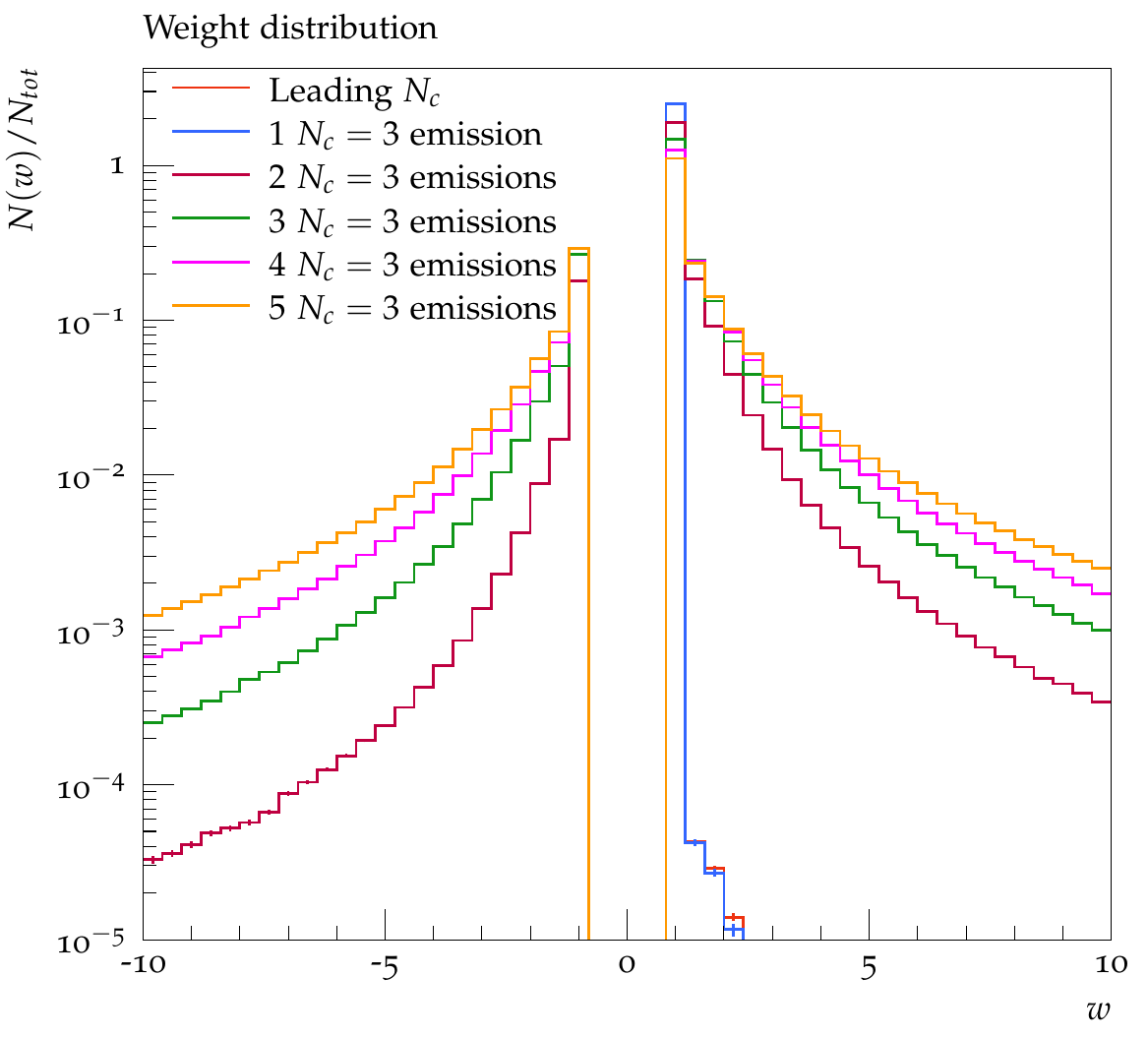}
  \includegraphics[width=0.45\textwidth]{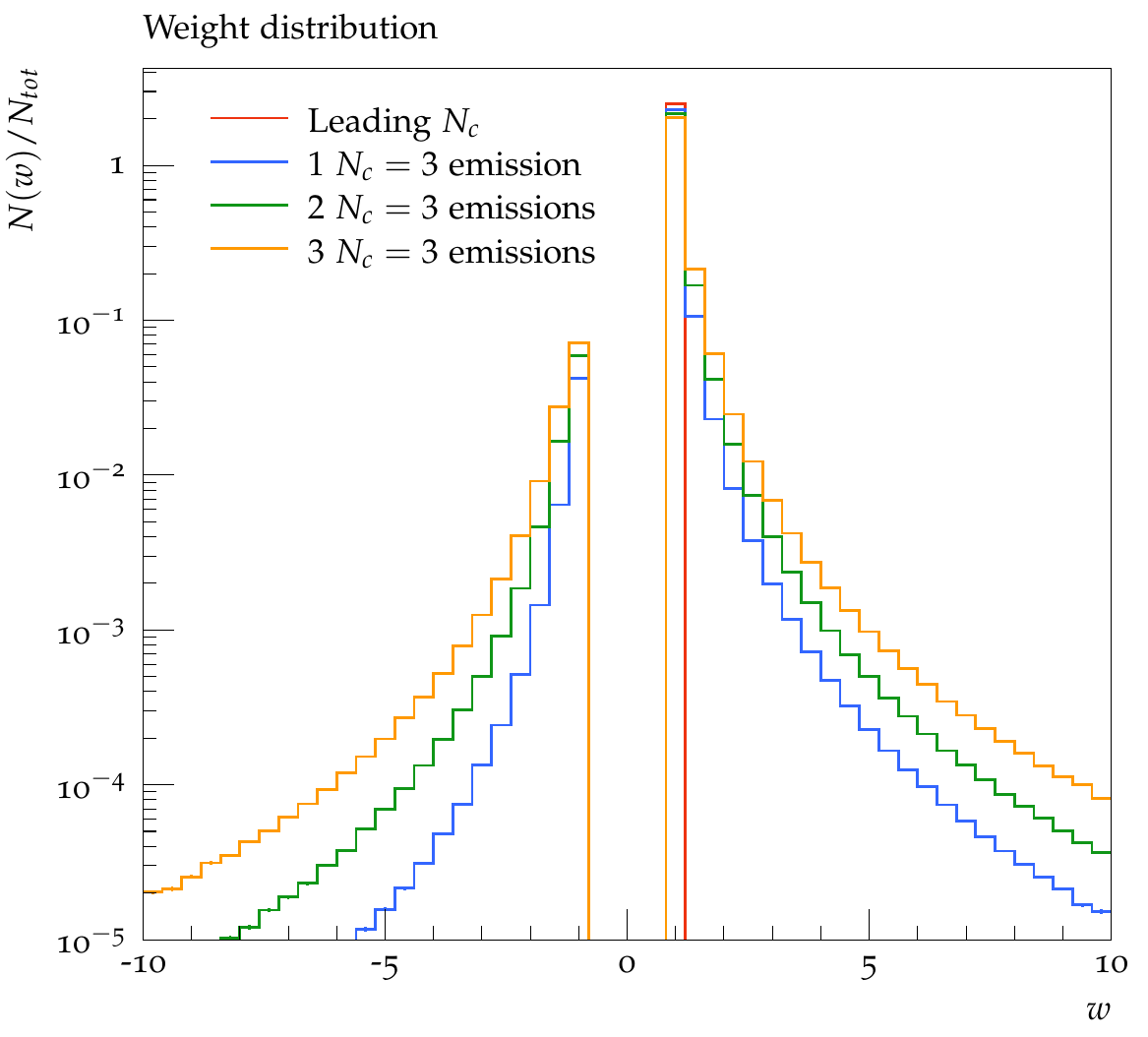}
  \caption{Weight distribution for $e^+e^-$ (left) and $pp$
    collisions (right) depending on the number of $\Nc=3$
    emissions allowed.
    All generated events are used in these plots, i.e., no further analysis
    cut is applied.
  } \label{fig:weights}
\end{figure}
In \figref{fig:weights} we show the weight distribution
for up to five/three subleading $\Nc$ emissions at LEP/LHC
respectively corresponding to up to six/seven $\qqbar$-pairs plus
gluons in the color bases. We note that although the weight distributions
get broader with the number of emissions, it stays
sufficiently narrow to ensure convergence for the considered
number of subleading emissions.
The number of colored partons, which we can practically include,
is therefore limited by the evaluation of scalar products
in color space, as described in \secref{sec:color}.
(We will find, however, that we are able to keep sufficiently
many full color emissions for standard hard observables to converge.)

Somewhat against intuition, we see a broader weight distribution
for LEP events than for LHC events, despite the fact that we
tend to have more colored partons at the LHC.
This can be attributed to the fact that the corrections
often tend to be negative at LEP (starting from $e^+e^- \to \qqbar$),
due to the negative contribution from coherent emission from the $\qqbar$-pair.
In line with this, we also note that
if we separately study $qq\to qq$, $qg\to qg$ and $gg\to gg$,
we find the largest weight variations for $qq\to qq$, another
case where we can expect large negative corrections from $\qqbar$-pairs.
\begin{figure}
  \centering
  \includegraphics[width=0.45\textwidth]{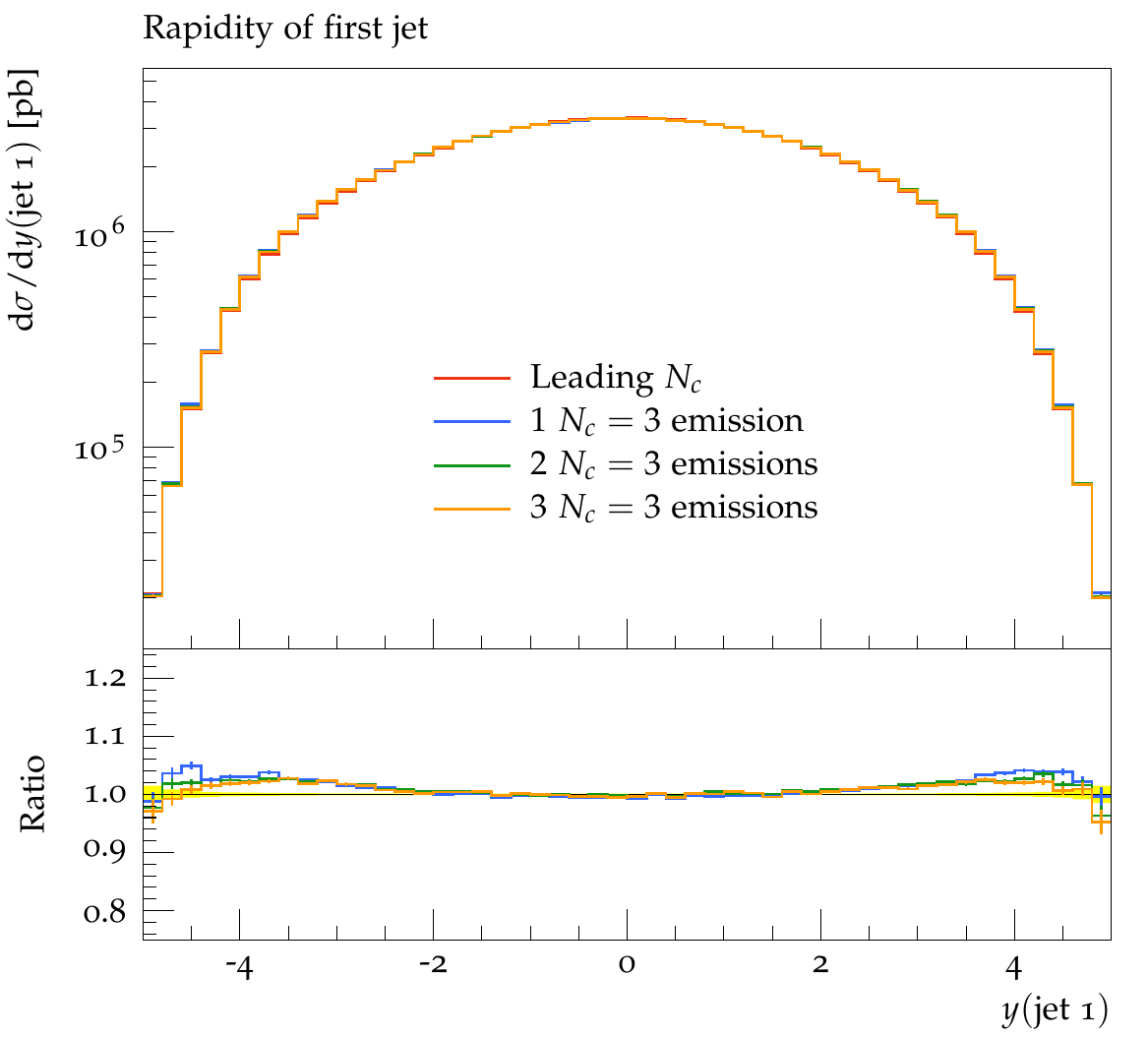}
  \includegraphics[width=0.45\textwidth]{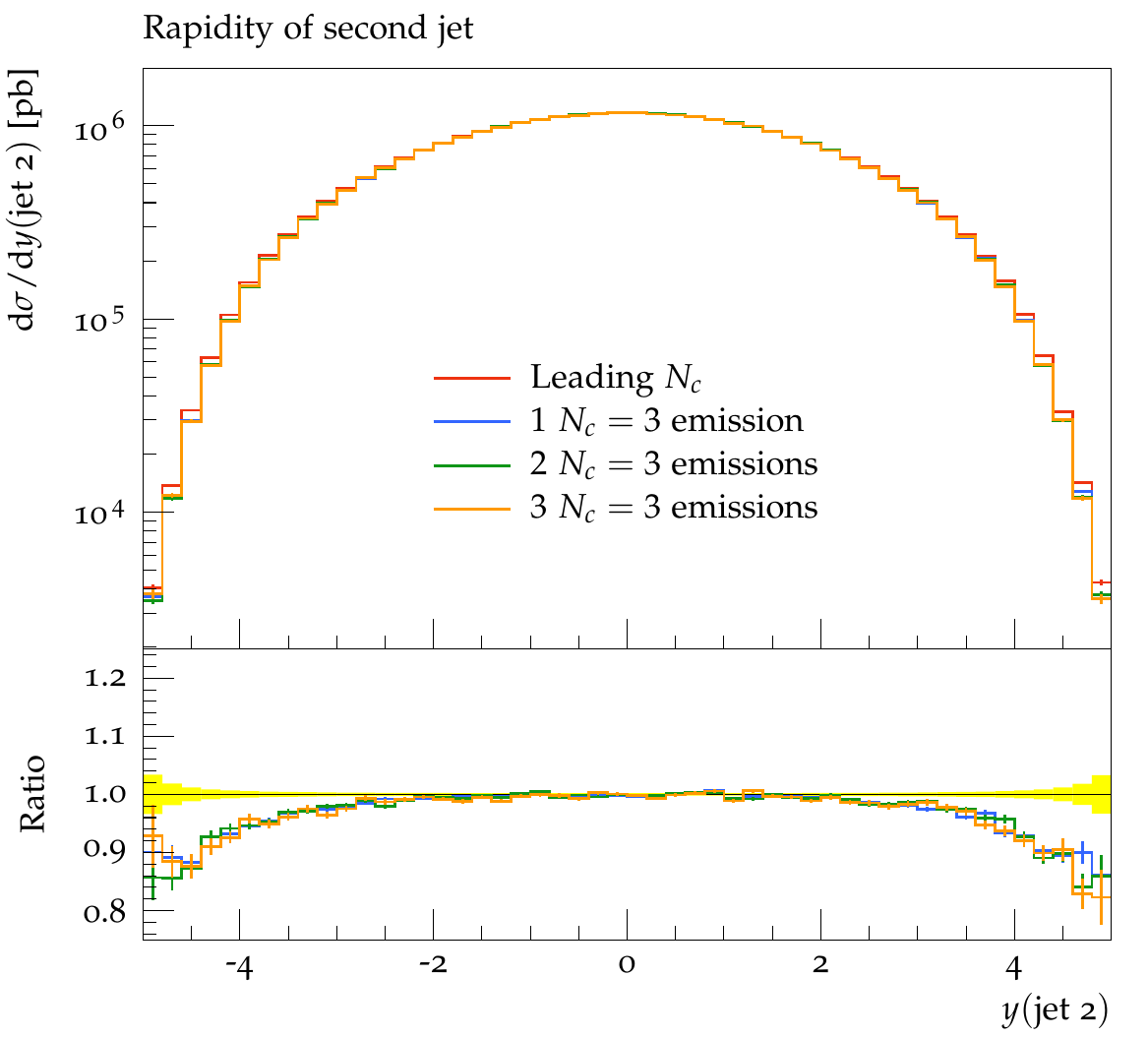}
  \caption{
    Rapidity distribution of the first and second jet as
    zero, one, two and three $\Nc=3$ emissions are kept.
  } \label{fig:rap}
\end{figure}

We next turn to the convergence of standard observables
with respect to the number of subleading $\Nc$ corrected
emissions.
As an example, we consider the rapidities of the hardest  LHC jets in
\figref{fig:rap}. As can be seen, the curves
converge as more subleading emissions are added, and in this
respect we see the same pattern for all standard hard LHC
observables; they all converge when up to three subleading
emissions are added, i.e., starting with a $2\to2$ topology and
adding three subleading emissions (followed by leading $\Nc$ showering)
gives results very similar to adding just two subleading $\Nc$ emissions
(followed by leading $\Nc$ showering).
LEP observables show a similar convergence pattern.
Only when explicitly considering very many jets, does the convergence fail.
This strongly suggests that for standard hard observables, subleading
$\Nc$ corrections can be well approximated by color correcting the
first few emissions.

\begin{figure}
  \centering
  \includegraphics[width=0.3\textwidth]{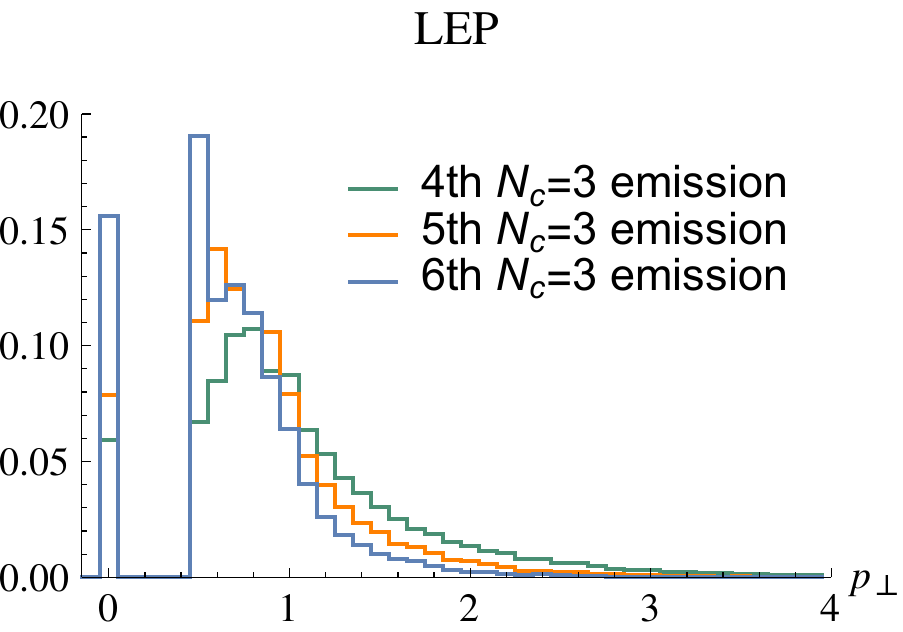}
  \includegraphics[width=0.3\textwidth]{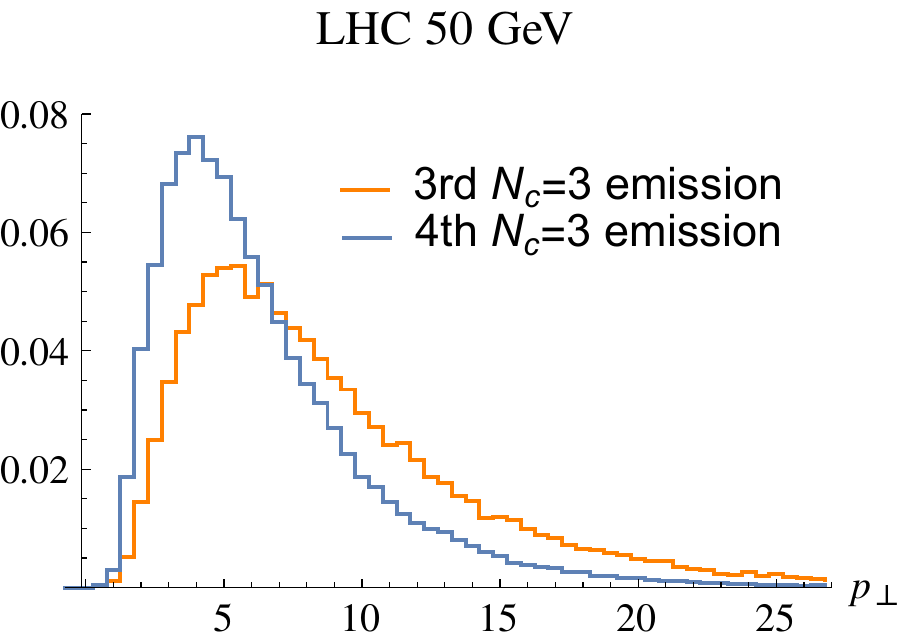}
  \includegraphics[width=0.3\textwidth]{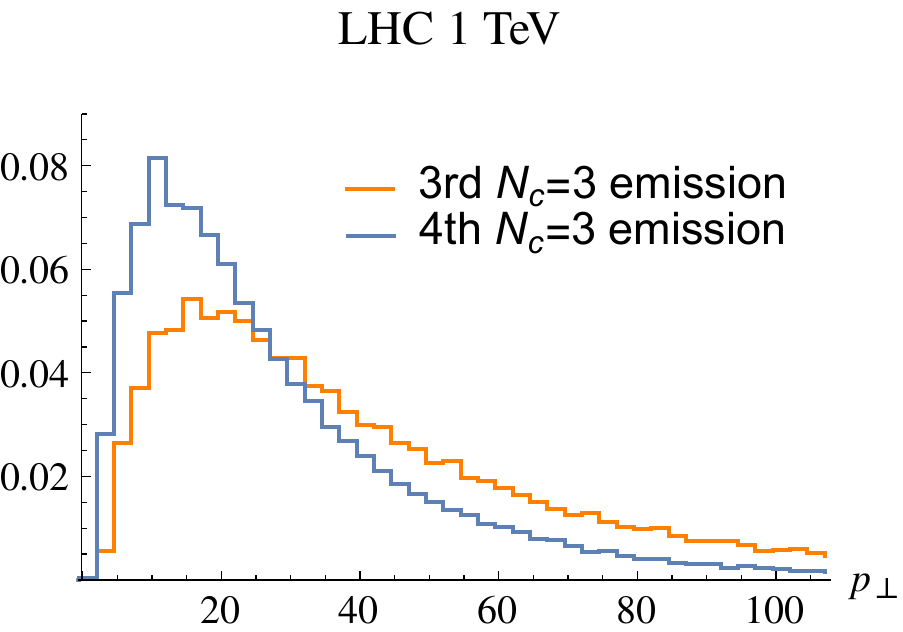}
  \caption{
    Transverse momentum distribution in the dipole frame of the last $\Nc=3$
    emission (orange) and the first non-corrected emission (blue).
    For LEP the second-to-last $\Nc=3$ emission is also shown (green).
    For LHC we use a 50 GeV (middle) and a 1 TeV (right) cut on the hardest parton.
    The shower cut-off at LEP is 0.45 GeV,
    and events which reach this scale
    are shown in the 0-bin above.
  } \label{fig:last_pT}
\end{figure}

The convergence can also be underpinned by studying the evolution
scale at which the $\Nc=3$ parton shower terminates, and
further evolution only is given by the leading $\Nc$ shower.  This is
investigated in \figref{fig:last_pT} where the transverse momentum, as
measured in the frame of the emitting dipole, is shown in
  orange for the last full color corrected emission, i.e., while
keeping up to three subleading emissions at the LHC and up to five
subleading emissions at LEP. For comparison we also show the distribution
of the first emission not to be corrected.

We find that the typical scale of the last color corrected emission is about
1 GeV at LEP and about 5-10 GeV at LHC, using a 50 GeV cut. Increasing
the cut to 1 TeV gives a much harder last subleading $\Nc$
corrected emission, as expected.

The convergence of observables is also in line with results from a
very recent paper on subleading $\Nc$ corrections at LEP \cite{Isaacson:2018zdi},
where the authors claim to observe good convergence while
keeping subleading corrections down to a variable cut-off
at around 3 GeV.

In  \cite{Isaacson:2018zdi} a Monte Carlo sampling over color is advocated,
and the point is made that this avoids the factorial scaling in
color space. In view of \figref{fig:last_pT}, we note, however, that
we can go further down in $p_\perp$ despite keeping the full color
structure. In fact, even if we limit ourselves to four subleading
$\Nc$ emissions, for which the time penalty due to color
structure treatment is negligible, we also go well below 3 GeV.
Therefore, in the case of hard observables at LEP, color
structure sampling seems to be of no benefit for the convergence of observables.

We have also performed a number of standard shower variation
checks, including shower scale variation and infrared cutoff variation.
While varying the shower scale, we find effects very similar to the
leading $\Nc$ case. 
Increasing the infrared cutoff from 1 to 2 GeV likewise results
in differences well comparable to the leading $\Nc$ shower, and so
does adding multiple interactions.

Finally, we have checked what happens if the shower is turned off completely
after one to three subleading emissions. Here, we find large
differences for the LHC, even if we keep up to three subleading $\Nc$
corrected emissions. We therefore conclude that it is essential
to keep showering beyond three color corrected emissions, but it is ---
for standard hard QCD observables, or likely any observable which is
mostly sensitive to the hardest jets --- not important to keep color
correcting the subsequent, softer and softer, emissions.
For observables depending on soft physics, the situation may be
different, as indicated below.

\subsection{LEP --- final state radiation}

\subsubsection{Parton level analyses}

\begin{figure}
  \centering
  \includegraphics[width=0.3\textwidth]{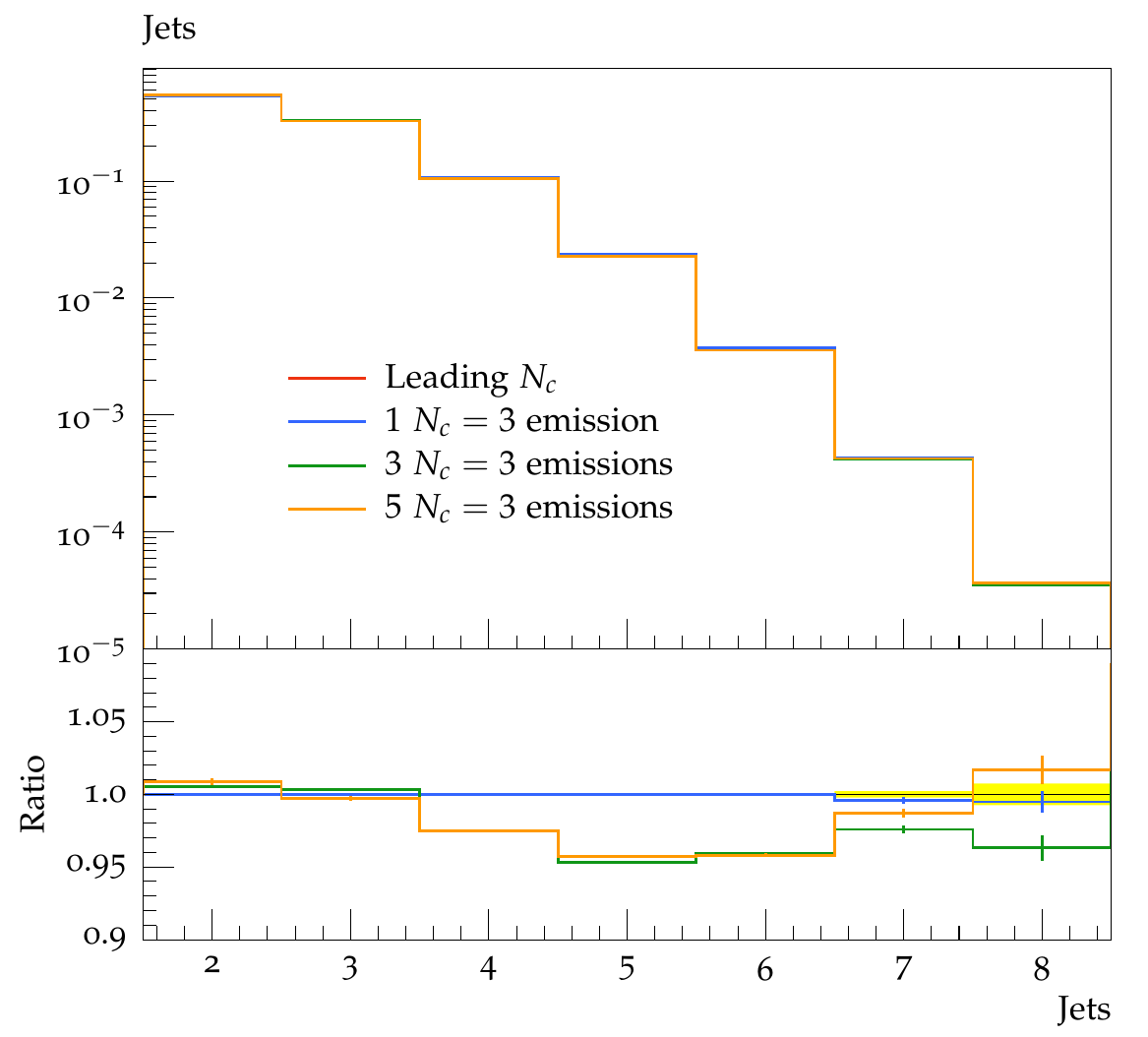}
  \includegraphics[width=0.3\textwidth]{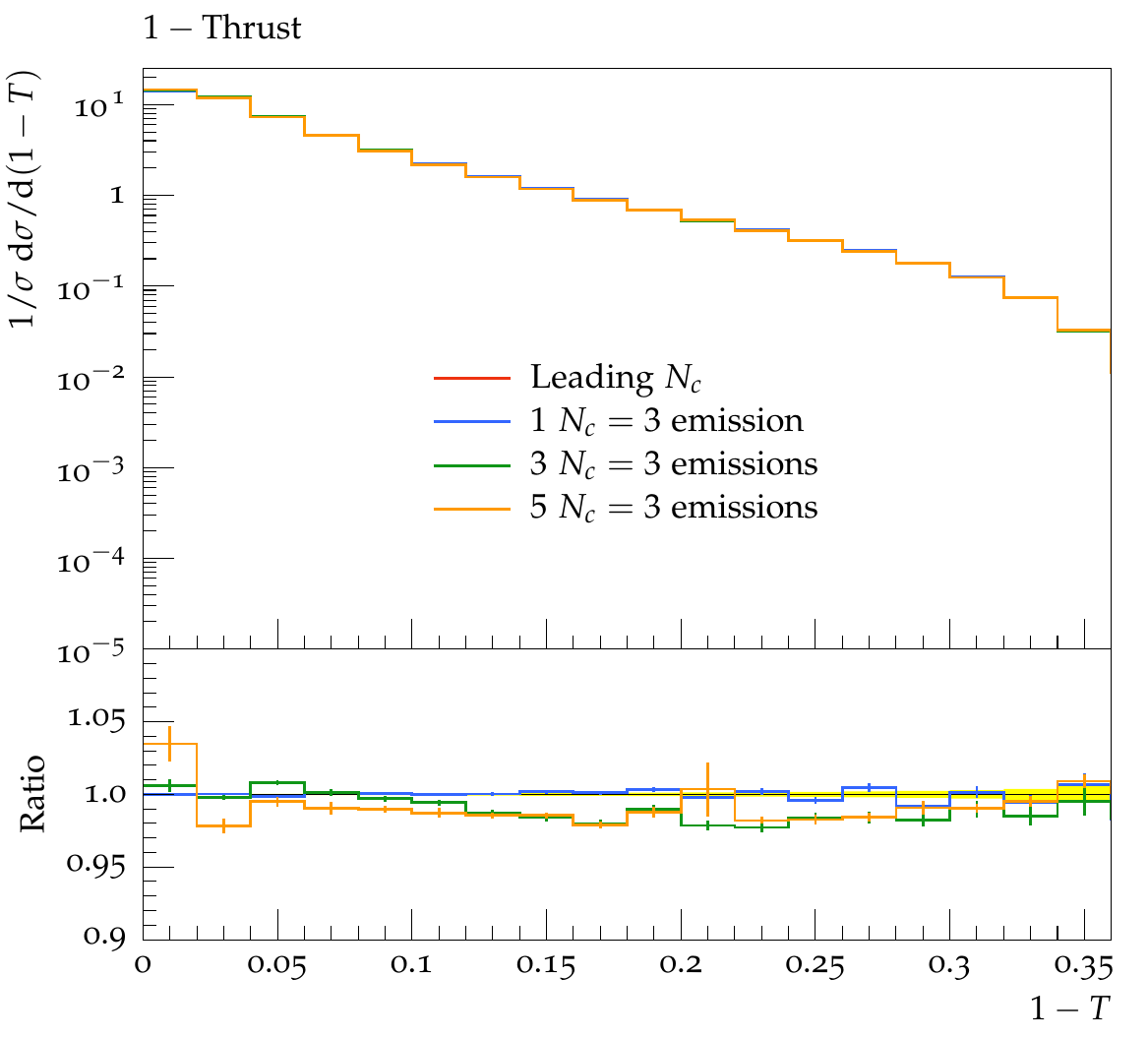}
  \includegraphics[width=0.3\textwidth]{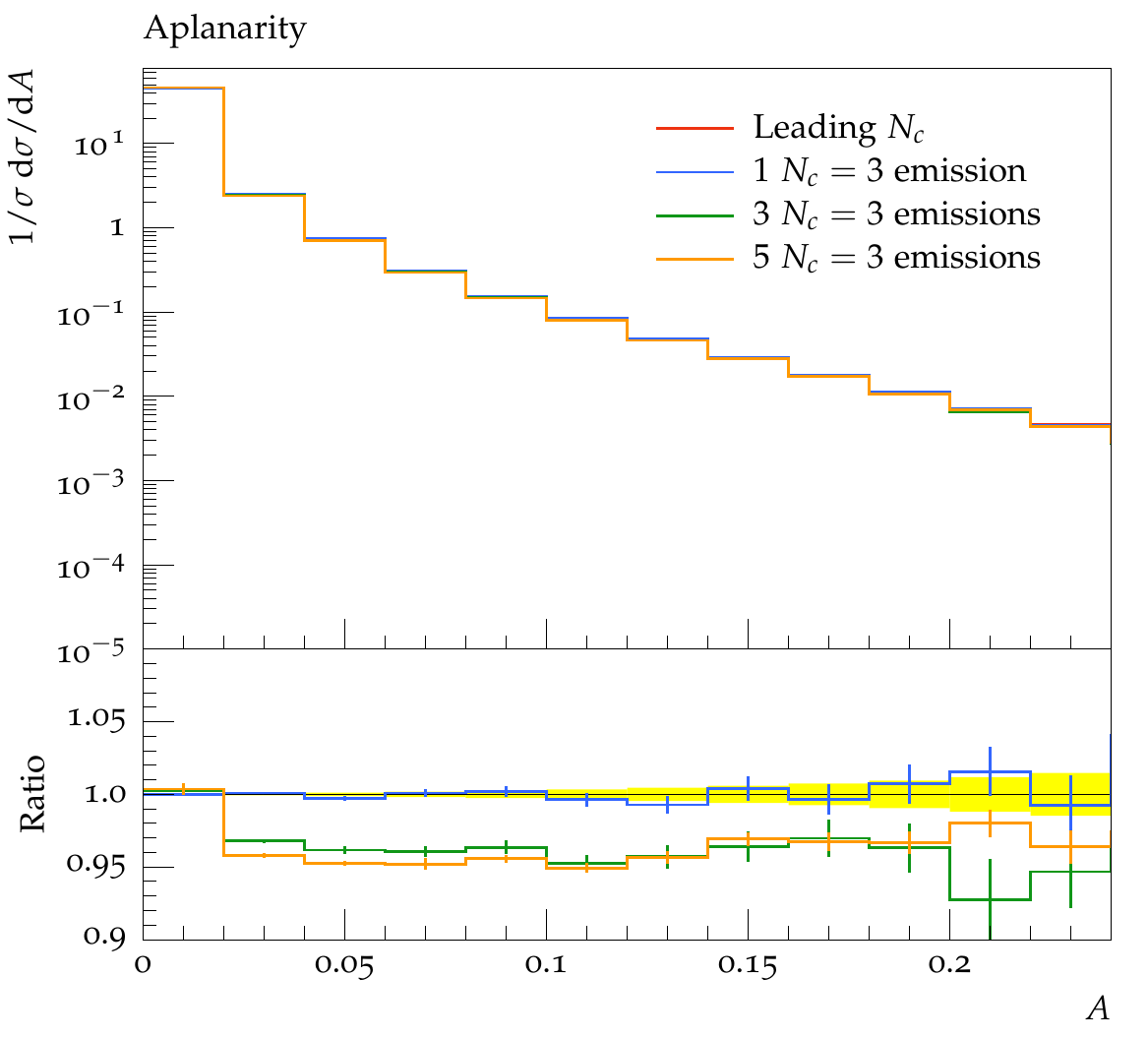}  
  \caption{
    Parton level plots for fraction of LEP-events containing $n$ jets with $E>5$ GeV (left),
    thrust (middle)
    and aplanarity (right),
    the jets have been clustered with the anti-$k_\perp$-type generalized 
    $e^+e^-$ clustering algorithm with $R=0.7$.
    For hadronized events the effect on thrust vanishes,
    whereas the effects on number of jets and aplanarity are somewhat reduced.
    Note that the case of one full $\Nc$ emission should agree with the
    leading $\Nc$ shower, as is seen.
  } \label{fig:ee}
\end{figure}

We first recapitulate the exercise from \cite{Platzer:2012np} and
run an $e^+e^-$ simulation at $\sqrt{s}=91.2$ GeV, without
hadronization, but this time including subsequent leading
$\Nc$ showering beyond the (up to) five subleading $\Nc$
emissions, as well as $g\to q\qbar$ splittings\footnote{The shower
  has also changed in several other respects compared to
  \cite{Platzer:2012np}, the momentum fraction integration boundaries
  have changed, the leading $\Nc$ assignment has been updated as
    described in \secref{sec:dipole}
  (having a large effect on predictions, similar to in \cite{Bellm:2018wwz}),
  the running of $\as$ is different,
  and we use the Sudakov veto algorithm from \secref{sec:sudakov}.
  For all these reasons, a direct comparison is not possible.}.
Again we find that the corrections to most LEP observables
are small. As examples of observables which show some effect,
we show the fraction of events containing $n$ jets with $E>5$ GeV,
the thrust distribution, and the aplanarity in \figref{fig:ee},
in all cases showing corrections below 10\%.
On the other hand, effects can be significant in tailored situations.
For example, considering the average transverse momentum with
respect to a thrust axis defined by the three hardest partons,
we find corrections above 10\%. We caution, however, that without modification
this is not an observable.
\begin{figure}
  \centering
  \includegraphics[width=0.3\textwidth]{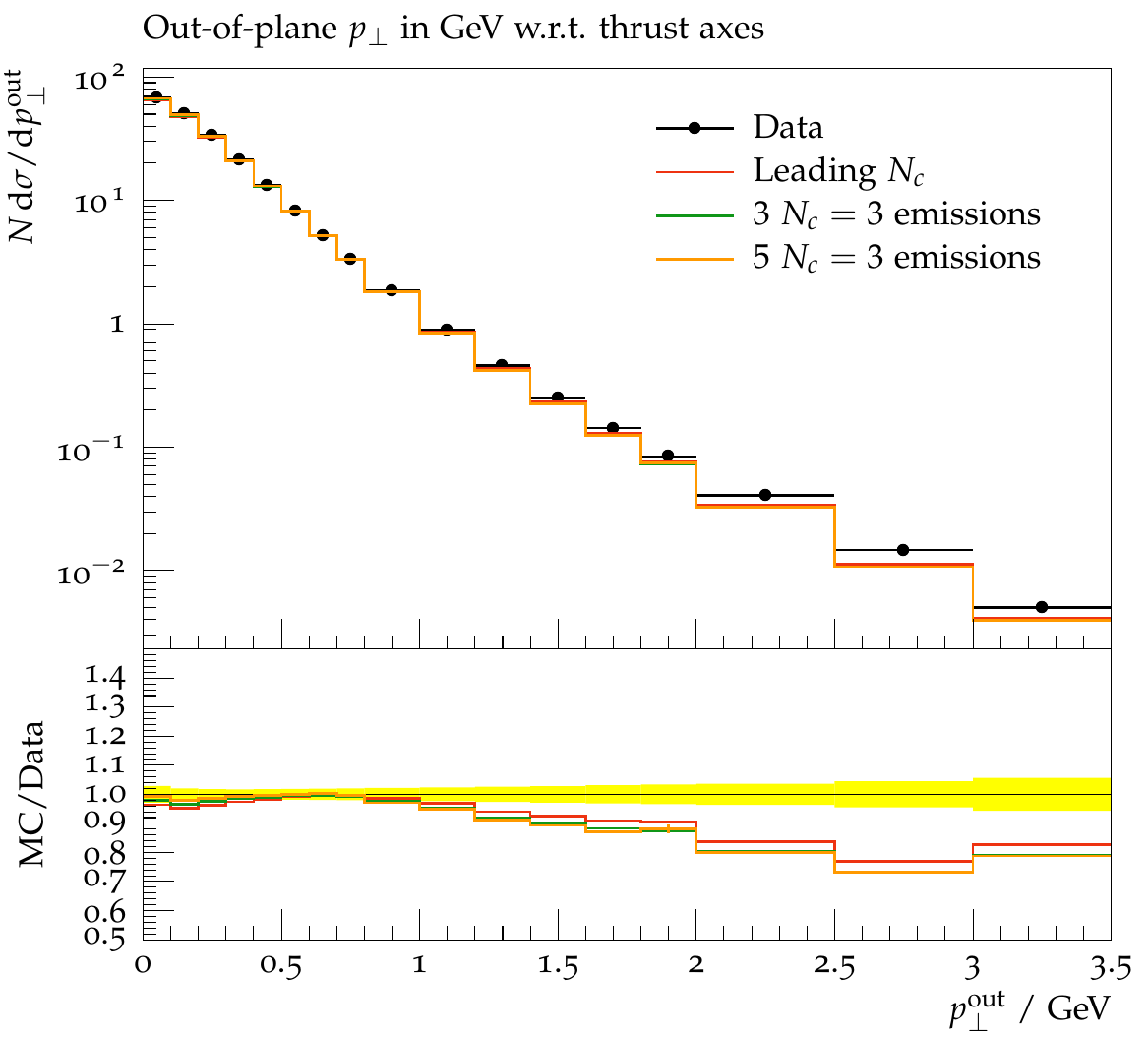}
  \includegraphics[width=0.3\textwidth]{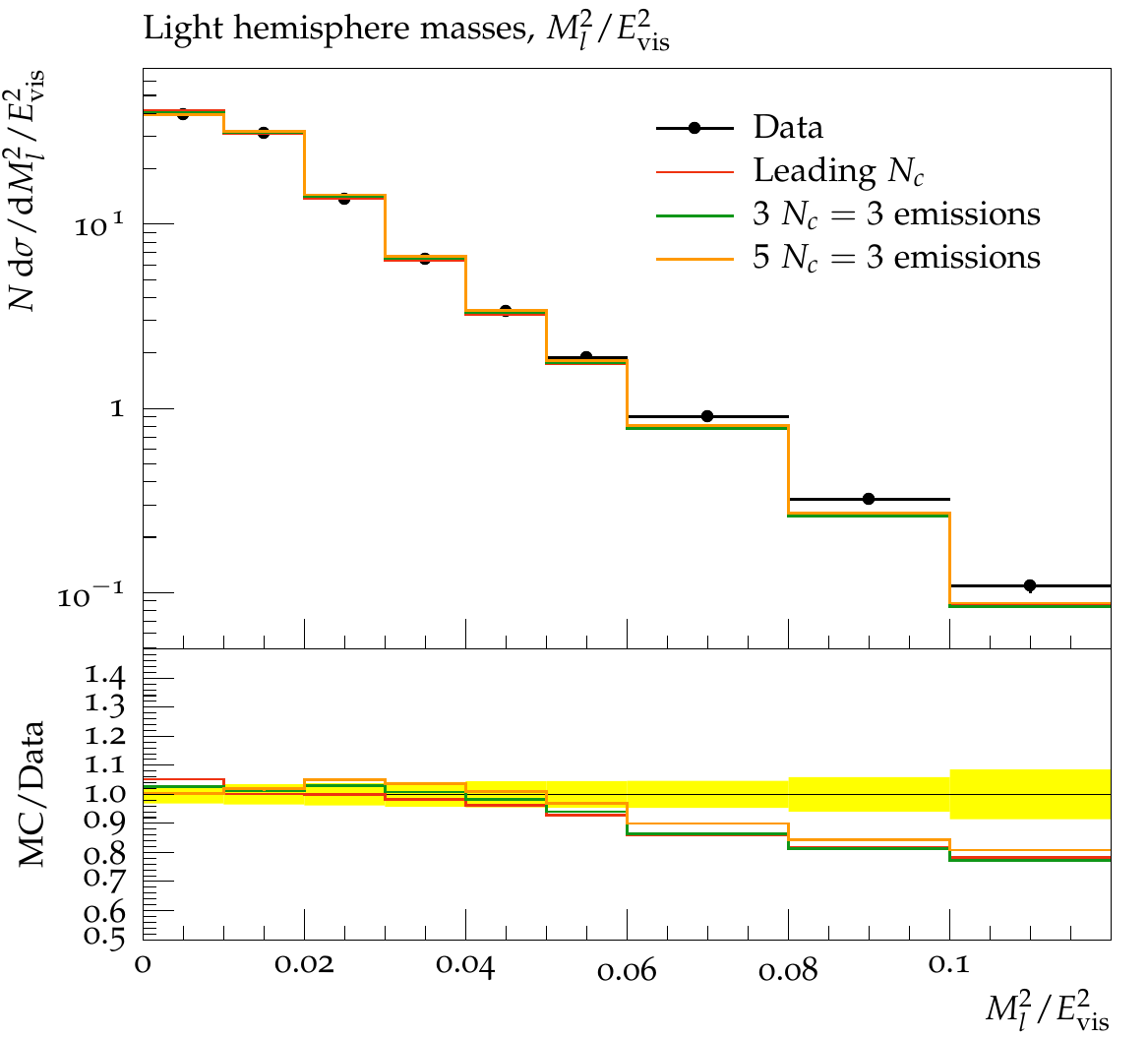}
  \includegraphics[width=0.3\textwidth]{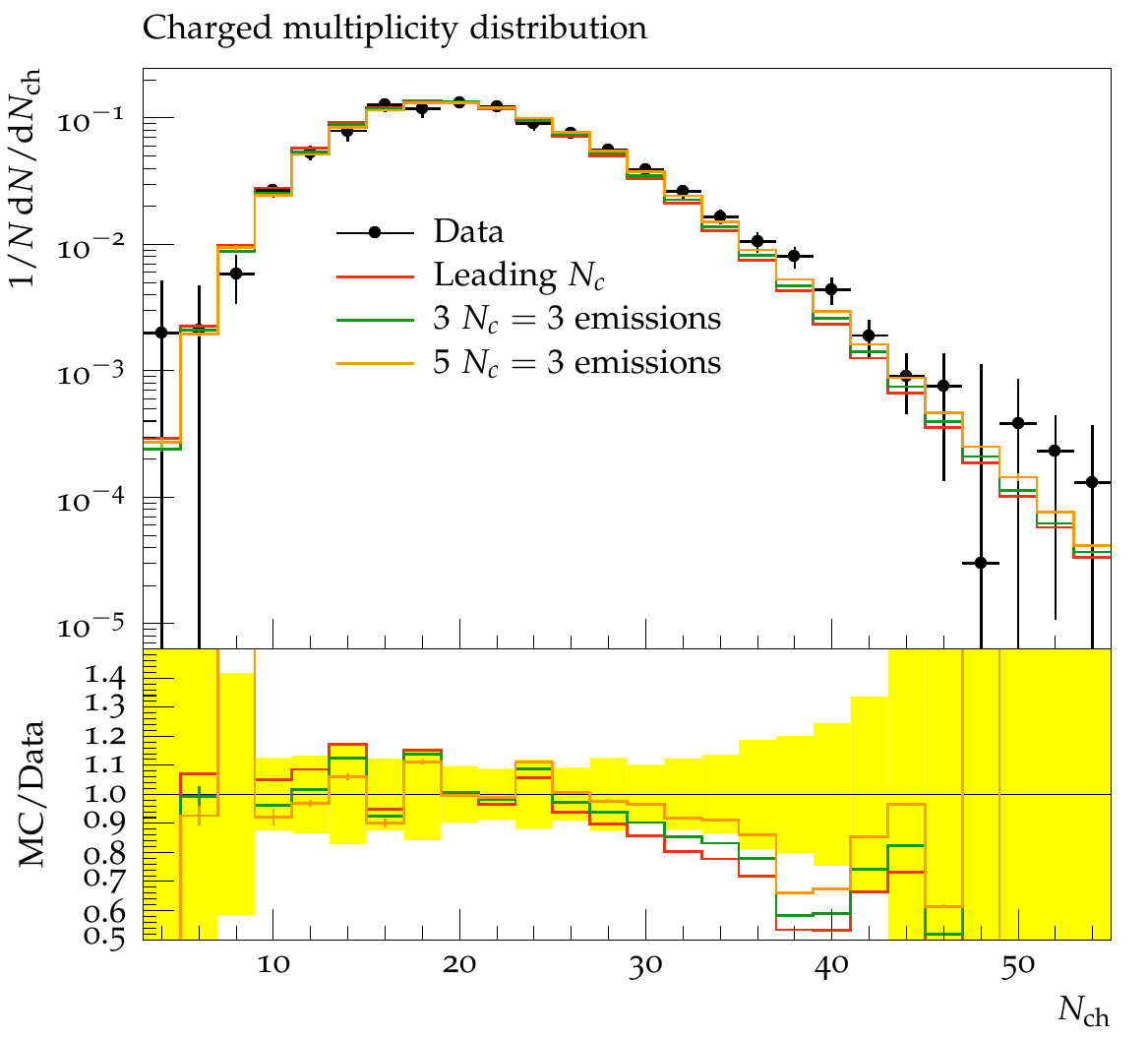}
  \caption{Out-of-plane $p_\perp$ w.r.t.\ the thrust and thrust major axes (left), light hemisphere mass (middle)
    and fraction of events containing $N_{\mathrm{ch}}$ charged particles, using to data and Rivet analyses from \cite{Abreu:1996na, Barate:1996fi}.
  } \label{fig:ee2}
\end{figure}

\subsubsection{Hadron level}

We have studied hadronized LEP events for a large class of
observables from \cite{Abreu:1996na, Barate:1996fi}.
For 
planarity,
sphericity,
oblateness,
and in- and out-of-plane $p_\perp$ w.r.t.\ sphericity axis,
we find small differences of a few percent or less, whereas
C-parameter shows some effects of 5-10\% in the low C-parameter
region.\
As examples, in \figref{fig:ee2}, we show the total out-of-plane $p_\perp$ (w.r.t.\
the plane defined by the thrust and thrust major axes)
and the light hemisphere mass in comparison to data from \cite{Abreu:1996na}.

In general
the deviation of simulated results compared to data from
\cite{Abreu:1996na,Barate:1996fi,Heister:2003aj} is clearly
dominated by other factors, and the overall description of
data does not change visibly.
Turning, on the other hand, to observables sensitive to soft
physics, we find larger differences. As an example we
show the charged multiplicity distribution in
\figref{fig:ee2} (right), compared to data from \cite{Barate:1996fi},
but we likewise see large effects for Durham jet resolution
variables in the soft region, on the color singlet cluster masses for
the Herwig hadronization model \cite{Bahr:2008pv} and on individual hadron
multiplicities.

While it is tempting to interpret the charged particle multiplicity
plot in \figref{fig:ee2}, as improved data description, we remark
that the hadronization model in use, is the standard Herwig cluster hadronization model
\cite{Bahr:2008pv}, with a leading $\Nc$ color flow, as described in \secref{sec:cmec}.
We therefore caution that the differences should only be
seen as an indication of effects on soft physics, from
the altered particle kinematics entering the hadronization.
While this can be interpreted as a need to retune the full color
parton shower, retuning shower parameters is beyond the
scope of the present paper.

\subsection{LHC --- coherent initial and final state radiation}

\subsubsection{Parton level analyses}

\begin{figure}
  \centering
  \includegraphics[width=0.45\textwidth]{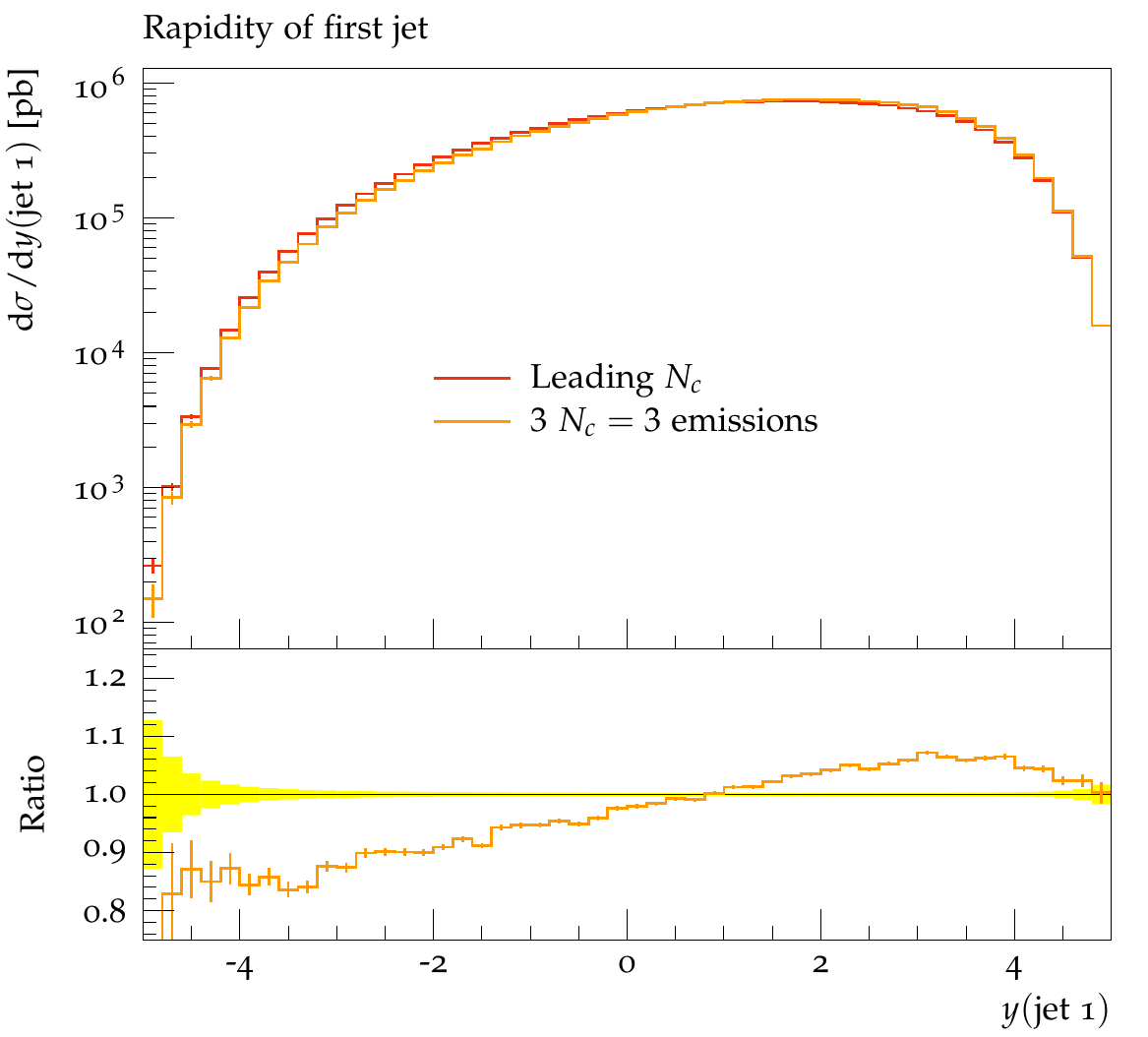}
  \includegraphics[width=0.45\textwidth]{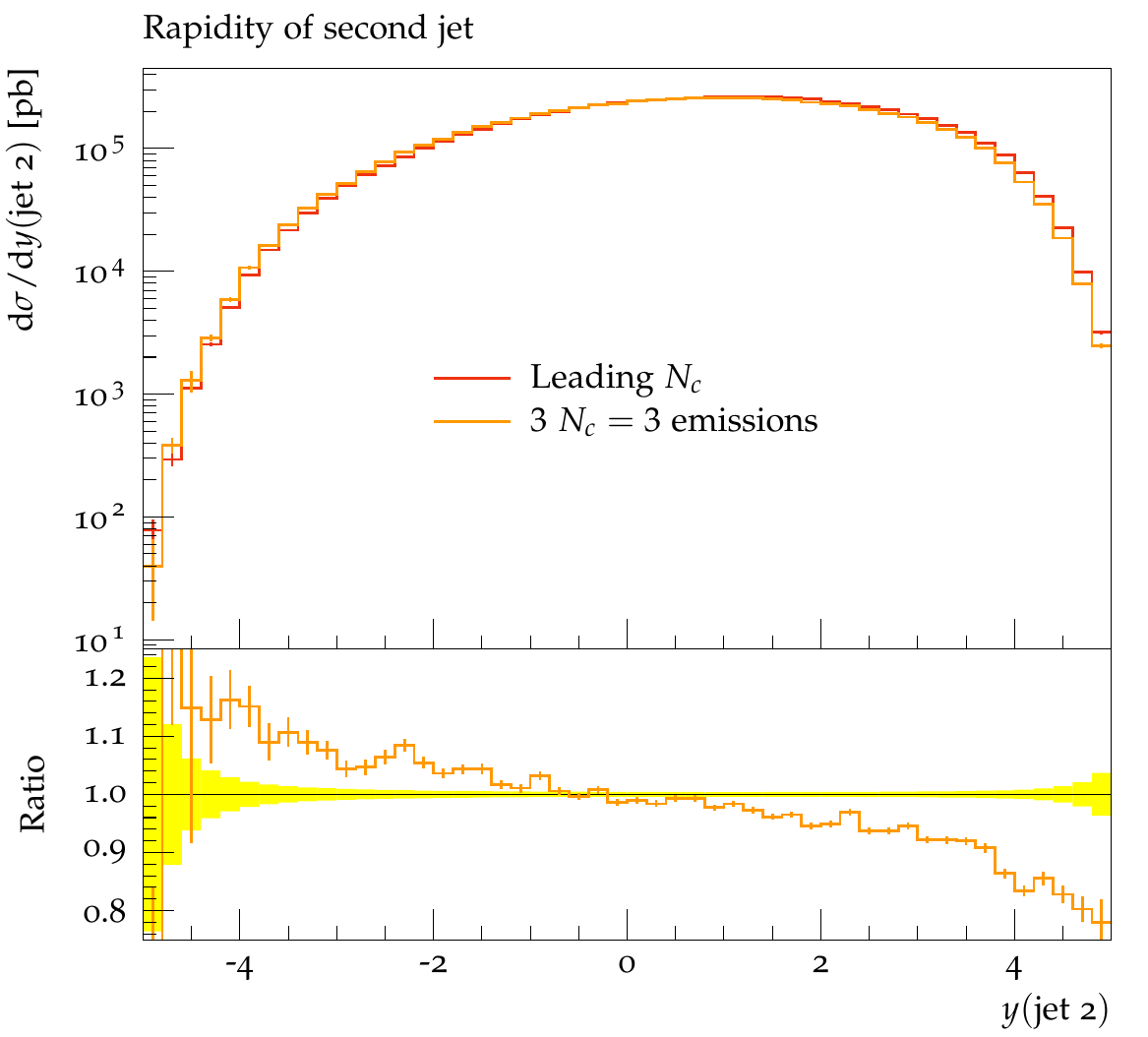}
  \caption{Rapidity distribution of the hardest and second hardest jet
    while considering only $qg\to qg$ scattering.
  } \label{fig:qg}
\end{figure}

We now turn to the LHC and start with reconsidering \figref{fig:rap}.
Considering the leading two jets, with our standard $p_\perp>50$ GeV analysis cut,
we see that they tend to have slightly different rapidity distributions,
with the second jet being more central.
For most other standard observables we find small differences,
of a few percent or less.
Nevertheless it is illustrative to separately consider scattering involving
different partons. Doing so we find that for $qq\to qq$ and $gg\to gg$,
the subleading $\Nc$ corrections are small for all studied observables,
whereas for $qg\to qg$, they can be more sizable.
In \figref{fig:qg} we therefore revisit the rapidity distributions of the two
leading jets, and find corrections going in opposite directions for the
two jets. Since $qg$-induced scattering contributes with a large fraction
of the cross section for the applied cuts, and since LHC data contains
$qg\to qg$ and $gq\to gq$ (along with all other processes), it can be concluded
that significant cancellation of subleading $\Nc$ corrections is present at LHC.

\begin{figure}
  \centering
  \includegraphics[width=0.3\textwidth]{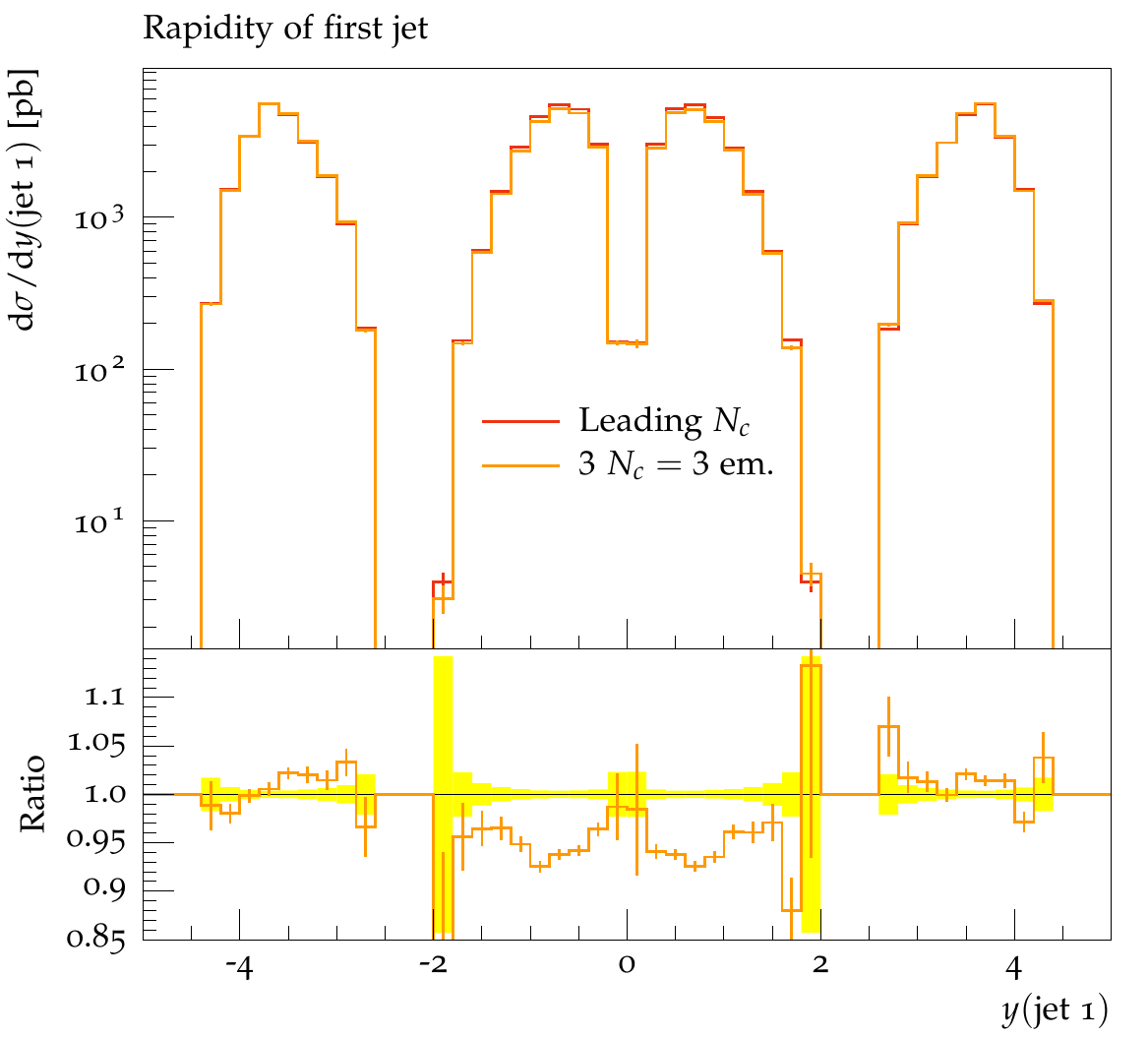}
  \includegraphics[width=0.3\textwidth]{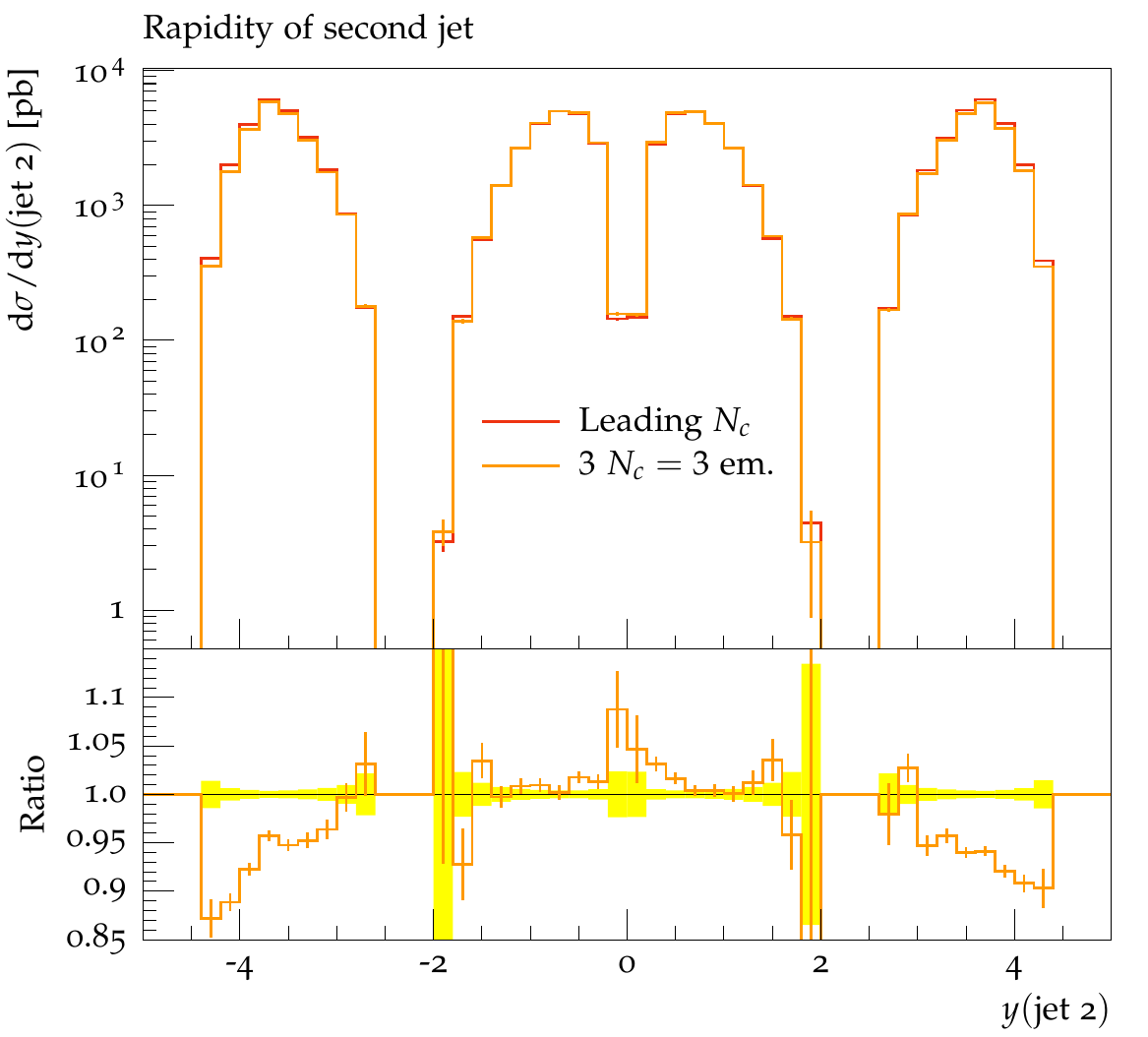}
  \includegraphics[width=0.3\textwidth]{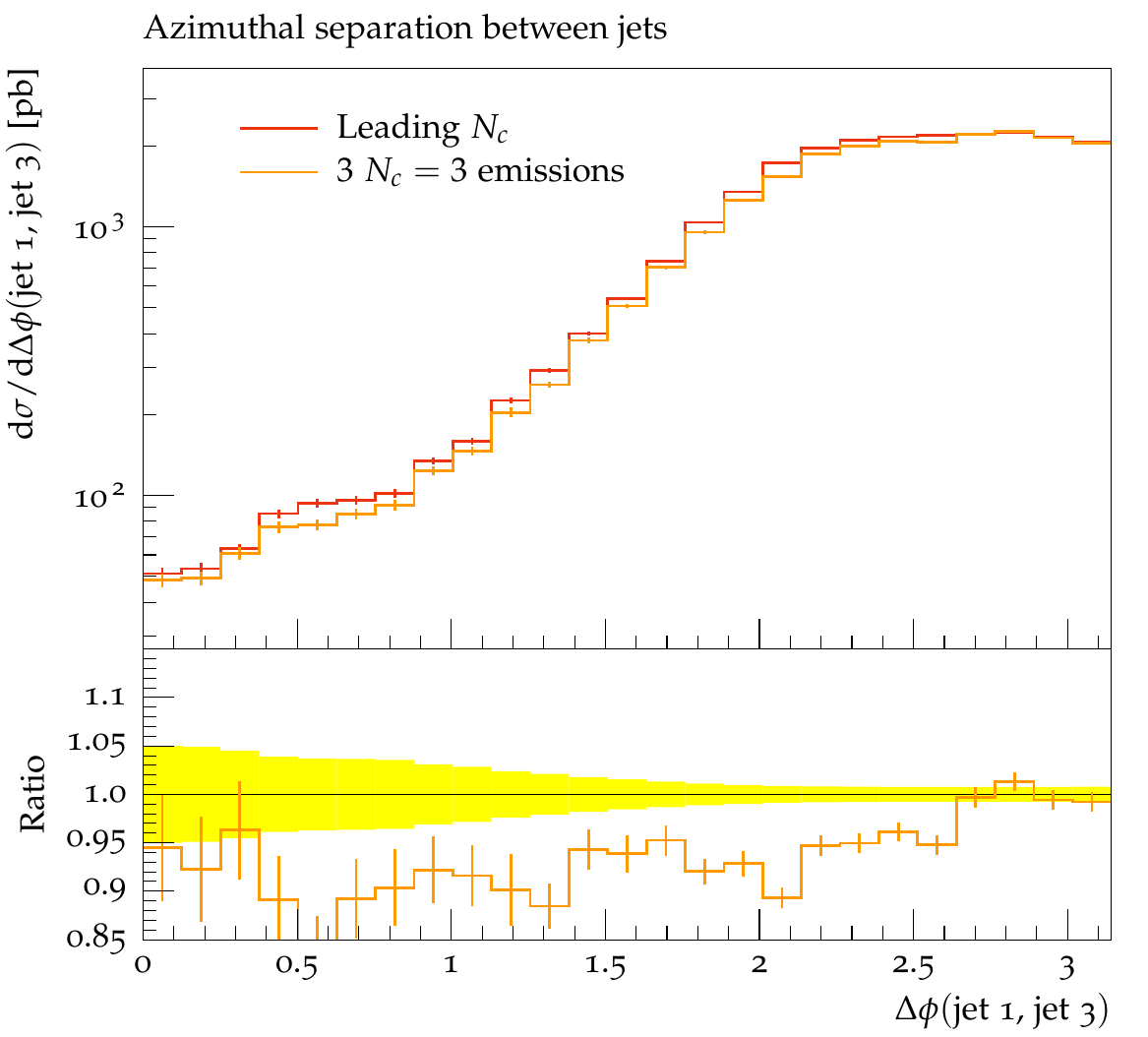}
  \caption{Rapidity distribution of the hardest jet (left), second hardest jet (middle)
    and separation in $\phi$ of the hardest and third hardest jets (right). Our standard analysis cut of $p_{\perp}>50$ GeV is used.
  } \label{fig:ppCF}
\end{figure}

Using the cuts of \cite{radek.simon}, 
\begin{align}
  400\text{ GeV} &< M_{12} < 600\text{ GeV}\ , \nonumber \\
  \label{eq:CF_cut_sumy}
  3.8 &< | y_1 + y_2 | < 5.2\ ,\\
  1.5 &< | y_2 - y_1 | < 3.5\ \nonumber ,
\end{align}
where $y_1$ and $y_2$ are the rapidities of the hardest and second
hardest jet and $M_{12}$ is the invariant mass of the two hardest jets,
events dominated by the hard process $qg\to{}qg$ can be statistically enhanced.
These cuts select events with one of the two hardest jets being
forward (in either direction) and one central.
With these cuts we find differences of $5-10\%$ for the rapidity
distributions of the hardest three jets, as illustrated in \figref{fig:ppCF}
for the hardest jet.
From \figref{fig:ppCF}, we see that in the $\Nc=3$
shower, the hardest jet tends to be central less often 
as compared
to the leading $\Nc$ shower. The rapidity distribution of the second
hardest jet 
shows that it is forward less often.
There are also $5-10\%$ differences in $\Delta\phi_{ij}=\phi_i-\phi_j$,
$\Delta\eta_{ij}=\eta_i-\eta_j$ and
$\Delta{}R_{ij}=\sqrt{\Delta\phi_{ij}^2+\Delta\eta_{ij}^2}$, for $i=1,2\ , j=3$.
As an example $\Delta{}\phi_{13}$
is also shown in \figref{fig:ppCF}. In general, with these cuts subleading $\Nc$ effects
show sizable corrections for many standard QCD observables.
\begin{figure}
  \centering
  \includegraphics[width=0.45\textwidth]{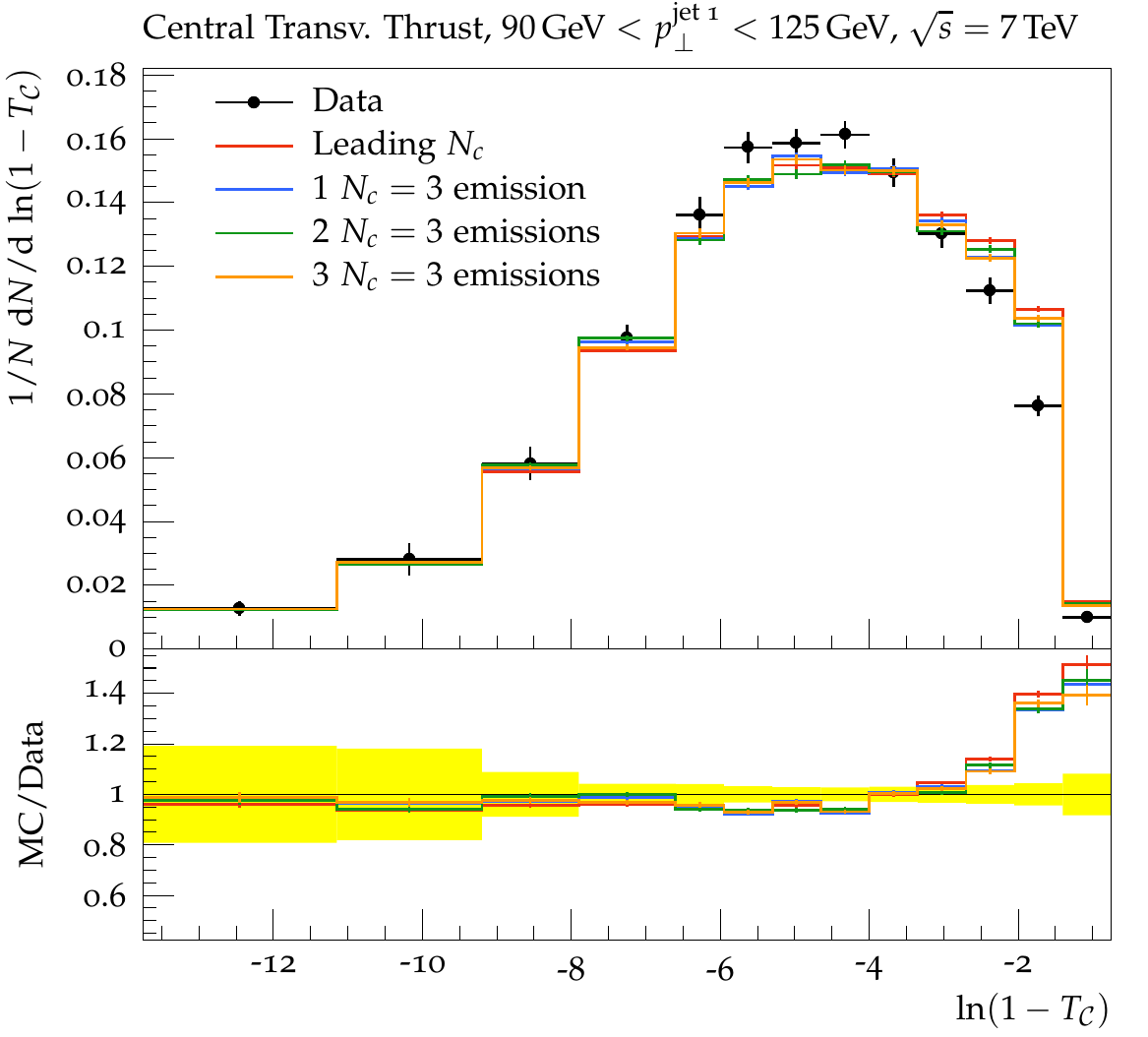}
  \includegraphics[width=0.45\textwidth]{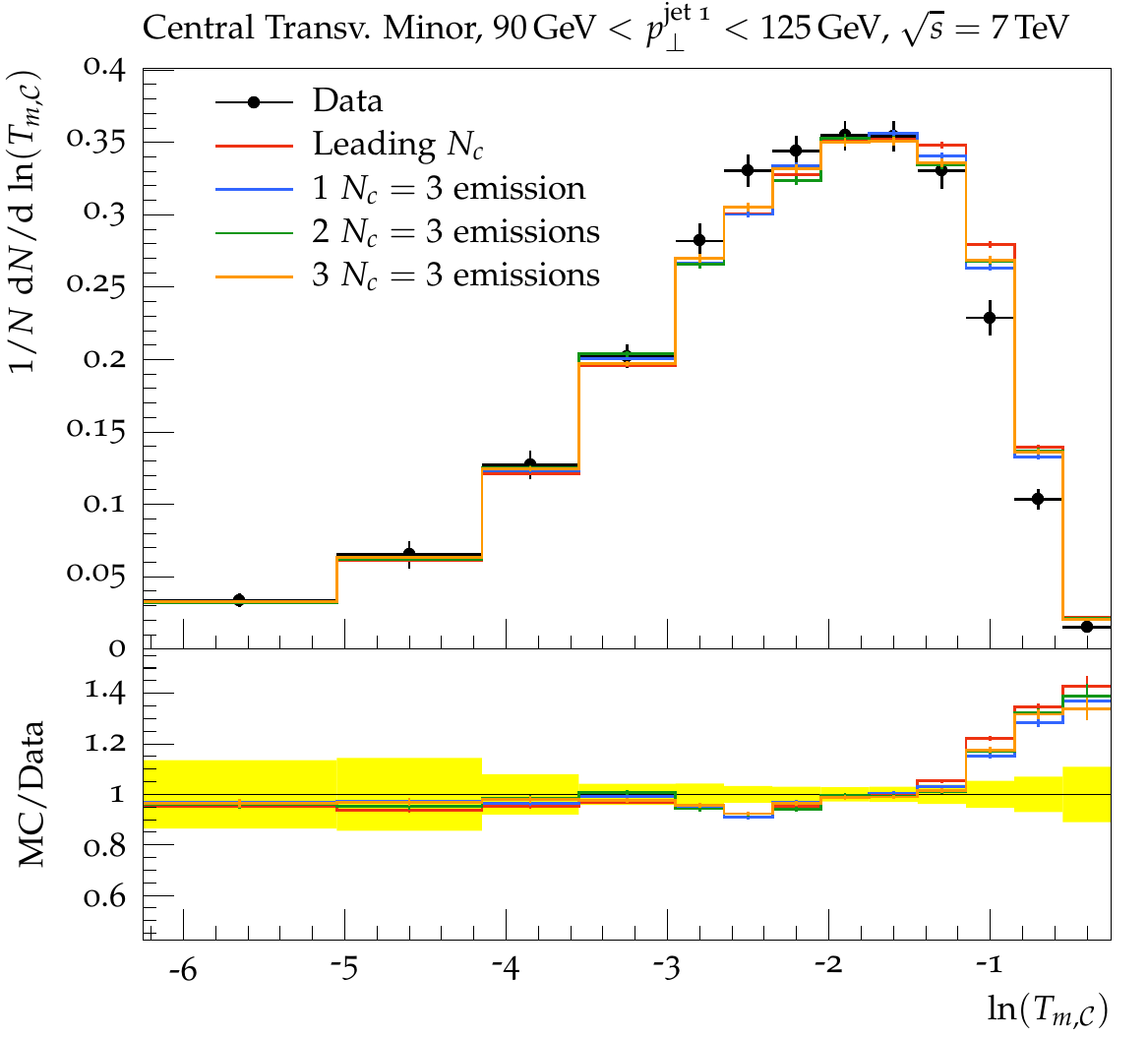}
  \caption{Distribution of $\log_e(1-T_C)$ (left) and $T_{m,c}$ (right) for $\sqrt{s}=7$GeV.
    Data and Rivet analysis are taken from \cite{Khachatryan:2011dx}.
  } \label{fig:thrust}
\end{figure}

\subsubsection{Hadron level analyses}

We now turn our attention to hadronized events and to comparisons with LHC data.
We have compared the subleading $\Nc$ corrected parton shower to experimental
data for a wide range of QCD observables, using data from
\cite{Khachatryan:2011dx,Aad:2012meb,ATLAS:2012al,Aad:2013fba,Chatrchyan:2013fha,Aad:2014pua,Aad:2016ria}.
  First, in \figref{fig:thrust}, we consider the event shape central transverse
  thrust, defined as \cite{Banfi:2010xy}
\begin{equation}
  T_{C}=
  \mbox{max}_{\hat{n}_T} \frac{\sum_i | \overline{p}_{\perp,i}\cdot \hat{n}_T | }{\sum_i p_{\perp,i}}\,,
\end{equation}
where $p_{\perp,i}$ is the transverse momentum of the central jet $i$, having pseudorapidity
$\eta < 1.3$, and $\hat{n}_T$ is the direction perpendicular to the beam axis,
which maximizes the sum. We also consider central thrust minor, defined again
in terms of central jets with $\eta<1.3$,
\begin{equation}
  T_{m,C}=
  \frac{\sum_i | \overline{p}_{\perp,i}\times \hat{n}_T | }{\sum_i p_{\perp,i}}\,,
\end{equation}
and compare to data from \cite{Khachatryan:2011dx}. As can
be seen we find relatively small corrections, at the 5\% level or below.

\begin{figure}
  \centering
  \includegraphics[width=0.45\textwidth]{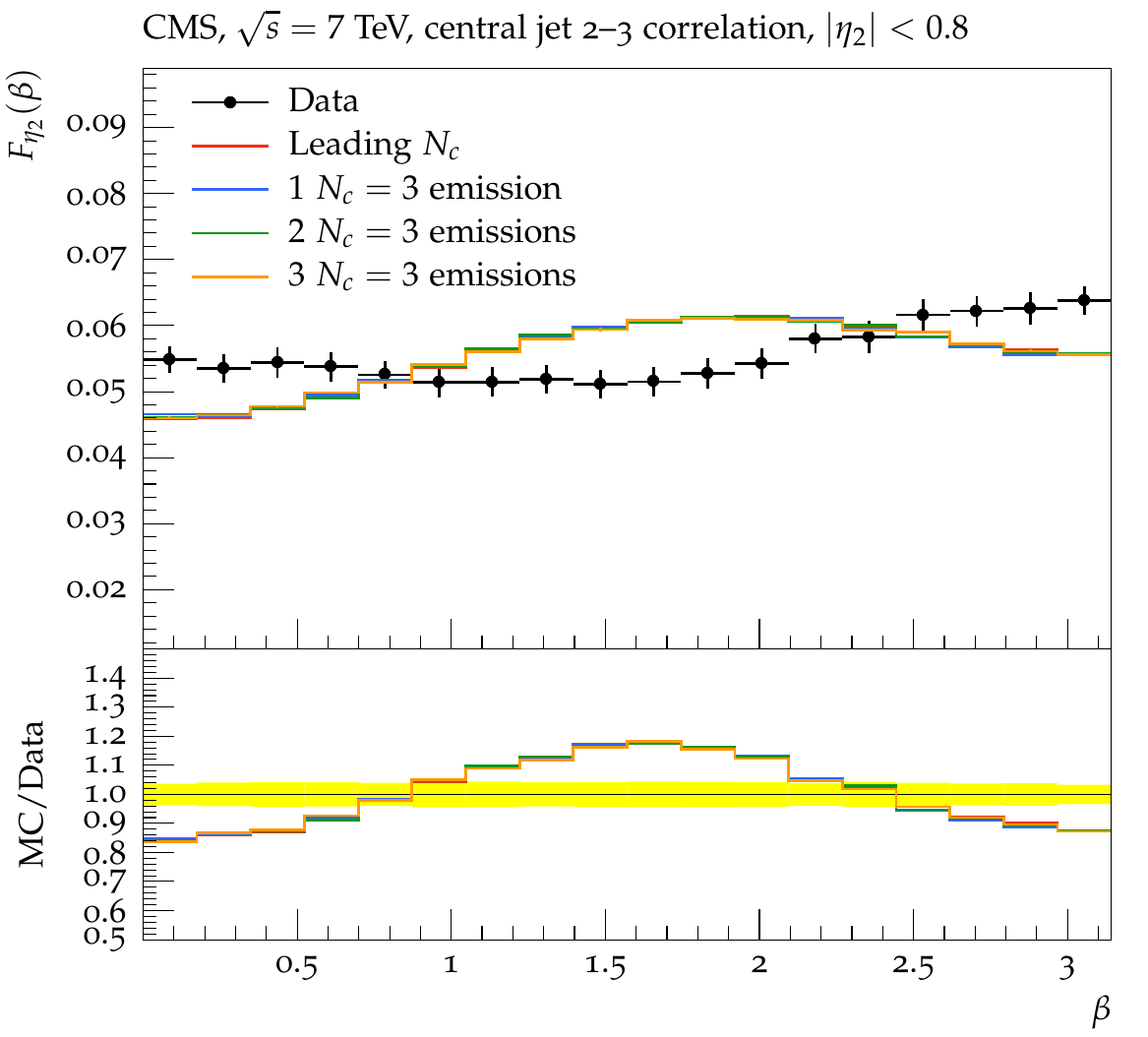}
  \includegraphics[width=0.45\textwidth]{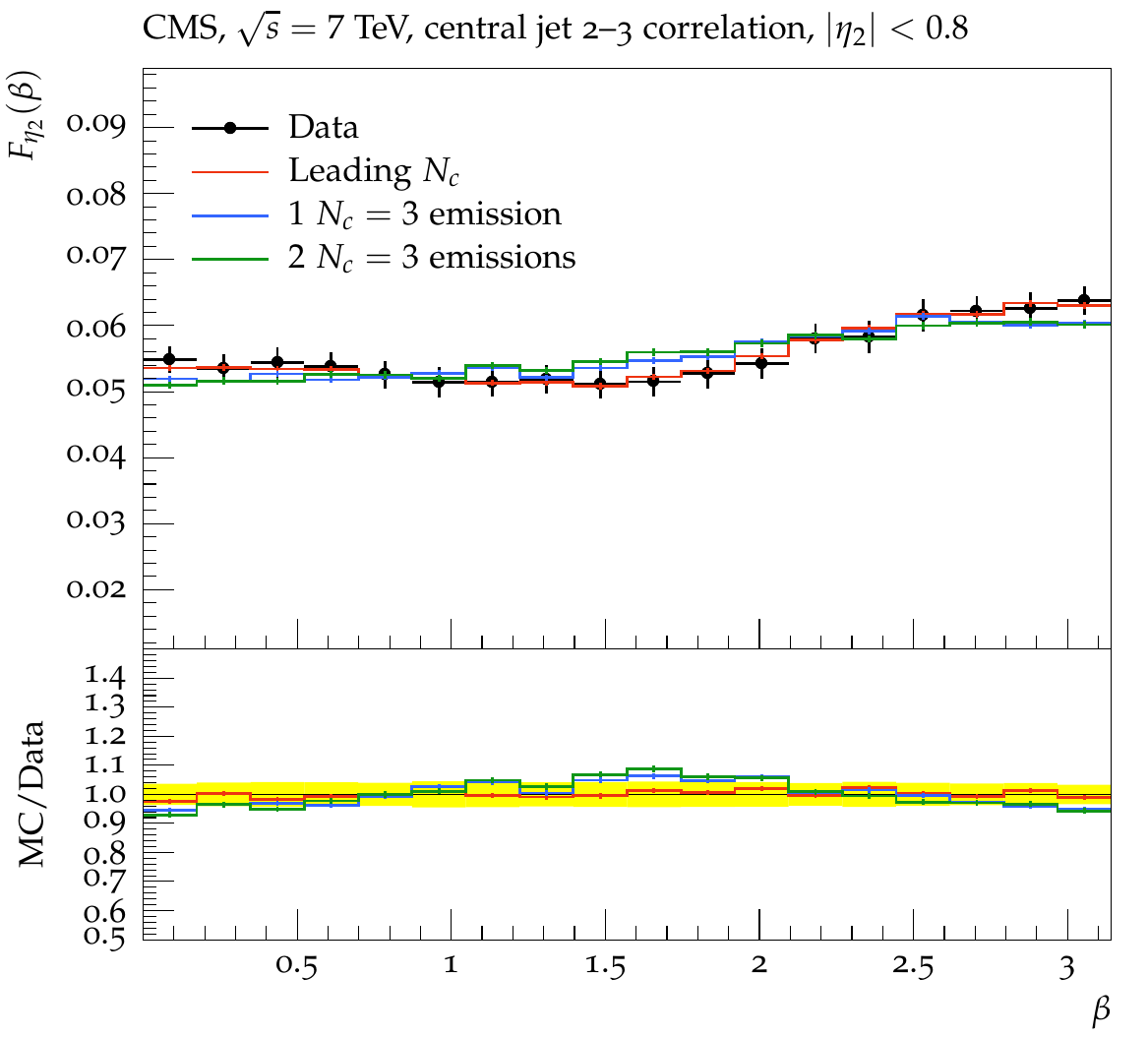}
  \caption{The angle $\beta$, defined as in \eqref{eq:beta} using (left)
    an underlying $2\to2$ hard process and (right) an underlying
    $2 \to 3$ hard process. Data and Rivet analysis are taken from \cite{Chatrchyan:2013fha}.
  } \label{fig:coherence}
\end{figure}

\begin{figure}
  \centering
  \includegraphics[width=0.45\textwidth]{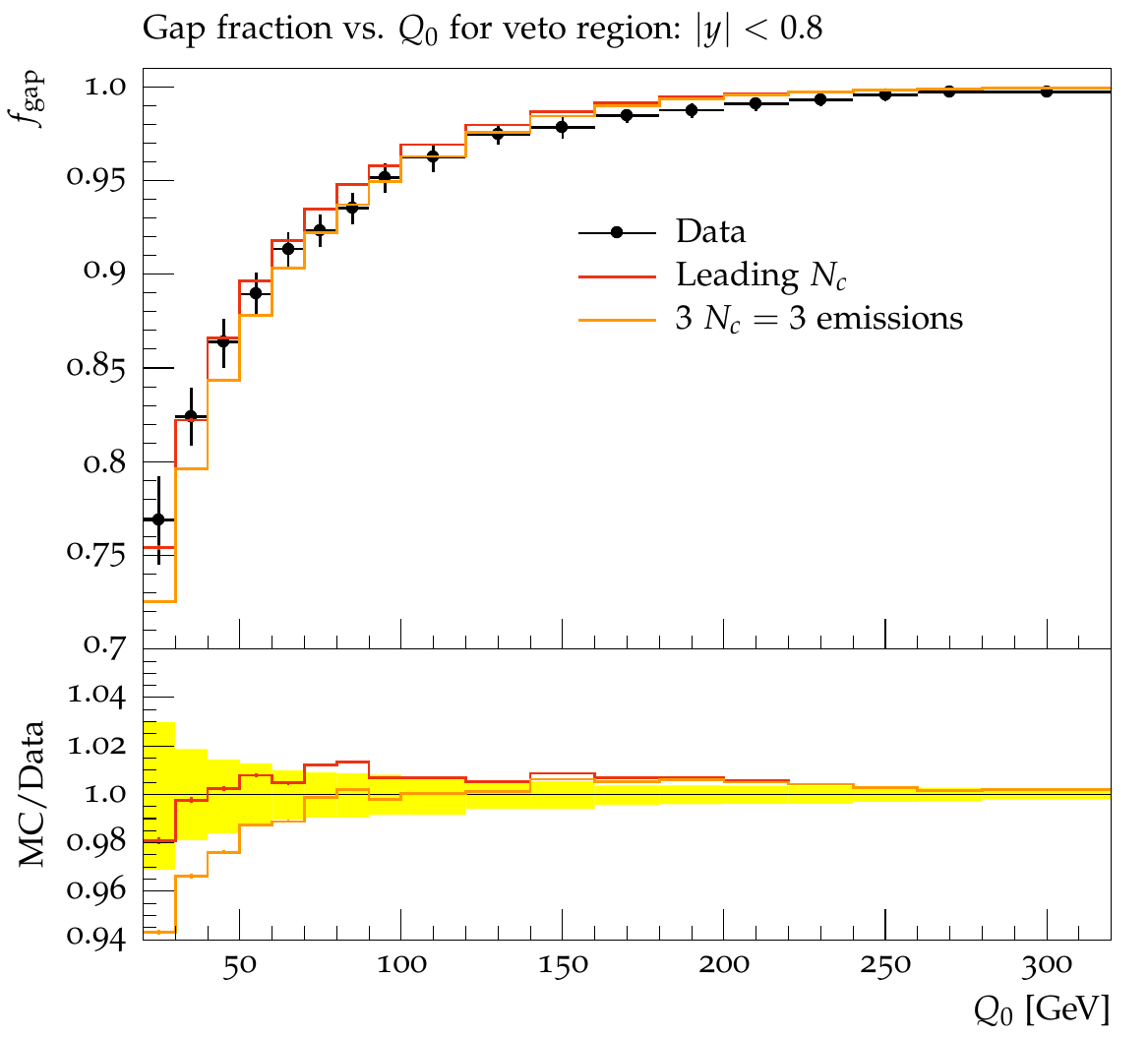}
  \includegraphics[width=0.45\textwidth]{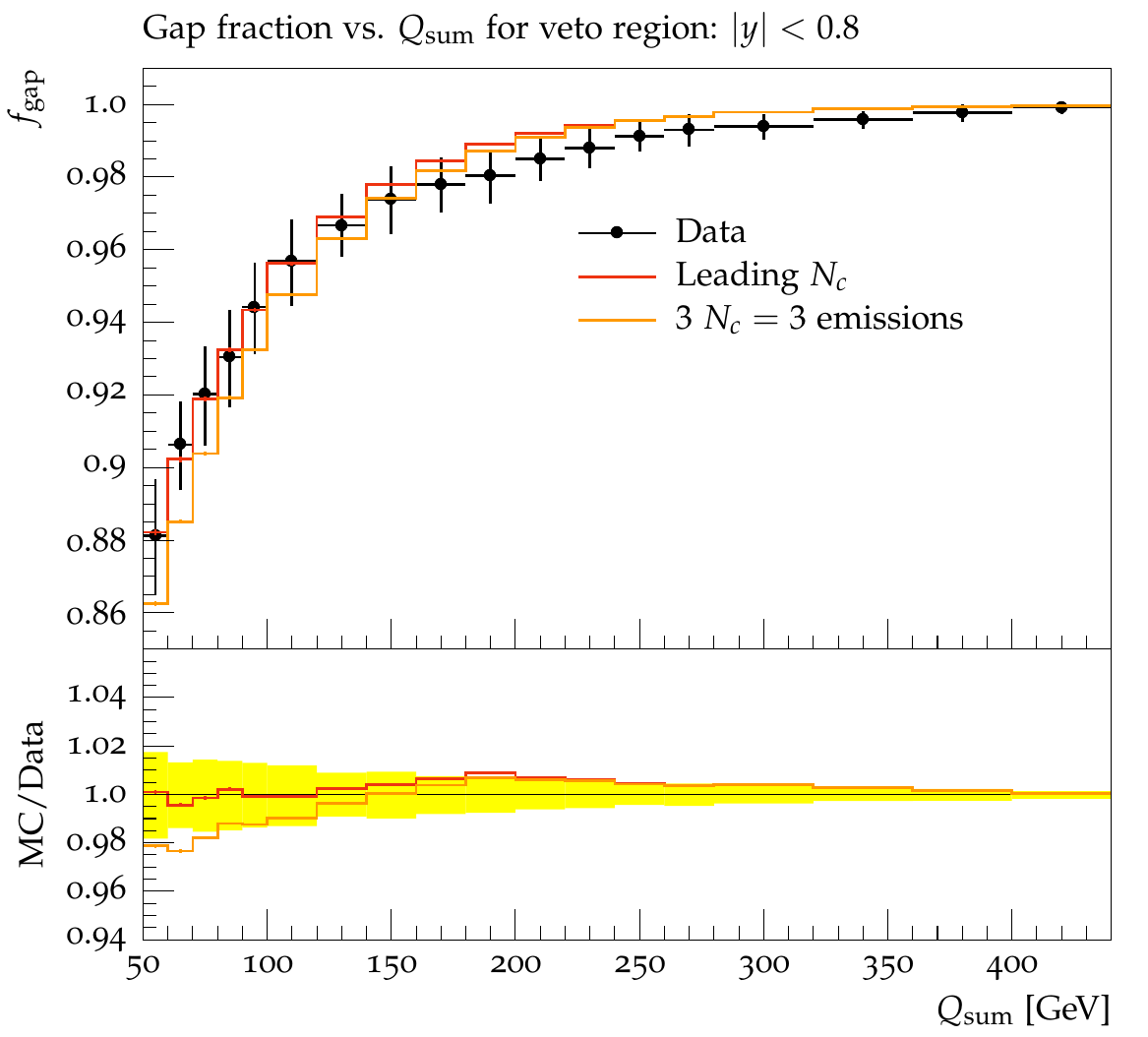}
  \caption{
      Fraction of events having no additional jet with $p_\perp$ above $Q_0$
      within a rapidity interval $|y|<0.8$
      (left) and fraction of events where the scalar sum of transverse momenta
      within $|y|<0.8$ does not exceed $Q_{\mathrm{sum}}$  (right) for
      $t\overline{t}$ events at $\sqrt{s}=7$ TeV.
      Data and Rivet analysis are taken from \cite{ATLAS:2012al}.
  } \label{fig:ttbar}
\end{figure}

A comparison to the jet shapes and jet masses for high $p_\perp$ jets,
from 
\cite{Aad:2012meb}, shows yet smaller subleading
$\Nc$ corrections, typically below a few percent. 

We have also compared data to the so-called color coherence effects from
\cite{Chatrchyan:2013fha}. In \figref{fig:coherence} we
show the distribution of the $\beta$-angle, defined in terms of the pseudorapidities
$\eta_2$ and $\eta_3$ and the azimuthal angles $\phi_2$ and $\phi_3$,
as
\begin{equation}
  \tan{\beta}=\frac{|\phi_3-\phi_2|}{\mbox{sign}(\eta_2)(\eta_3-\eta_2)}\,.
  \label{eq:beta}
\end{equation}
Experimental data is first compared to shower predictions using
a $2\to 2$ hard matrix element, and then using a $2\to 3$. We clearly see
that the use of a $2\to 3$ hard matrix
element significantly improves the description of data, whereas adding
subleading $\Nc$ showering to the $2\to2$ process changes the distribution
compared to the leading $\Nc$ shower very marginally.
This casts doubt upon the description of this observable as probing
color coherence, and rather illustrates its dependence on the hard
matrix element (or alternatively on other details of the shower algorithm).

We have also investigated the effects from subleading color contributions
on top-pair production, and compared to data from 
\cite{ATLAS:2012al, Aad:2013fba}.
Here we find that the jet shapes from \cite{Aad:2013fba} are
essentially unaltered, whereas the measurement of the additional jet
activity in $t\overline{t}$ events \cite{ATLAS:2012al}
show intriguing effects, displayed in \figref{fig:ttbar}. In particular
we note that the data description improves in the region of a modest
ratio between $Q_0/Q_{\mathrm{sum}}$ and the hard scale.
For gap observables, like in \figref{fig:ttbar}, with a large scale
hierarchy, we remark that we can expect effects from resummation
of virtual gluons, which we do not include in this paper.
These subleading $\Nc$ corrections may be sizable \cite{Forshaw:2007vb,Cox:2010ug,DuranDelgado:2011tp,Martinez:2018ffw}.

As a clean test of initial state radiation, we have also compared
the $\Nc=3$ shower to event shapes in leptonic $Z$ decay events,
\cite{Aad:2016ria}. Here, as expected, having fewer
colored particles in the hard process, we find very small corrections,
at the percent level or below.

\section{Conclusion and outlook}
\label{sec:conclusion}

In this paper we have investigated the effect of keeping the full color structure in parton
shower emissions in realistic simulations of LHC and LEP events.
This is pursued within the dipole shower of the Herwig 7.1 framework \cite{Bellm:2017bvx}
as color matrix element corrections.

The $\Nc=3$ color shower corrects
the first few (five for LEP and three for LHC) emissions
using the full $\Nc=3$ emission pattern.
All hard observables we have studied, with the exception of
observables explicitly considering very many jets, have
converged with respect to keeping additional subleading $\Nc$
color corrected emissions.
The convergence can also be underpinned by noting that the
last emission to be corrected, corresponds to a low 
evolution scale, in comparison to the studied observables,
see \figref{fig:last_pT}.
In particular for LEP, we typically go down to evolution
scales around one GeV.

Our results show that, for most QCD variables at the LHC, the subleading $\Nc$ effects are,
similarly to at LEP, of the order of a few percent. At the LHC we can, however, see larger effects, $10-20\%$, for the tails of the rapidity distribution of the second hardest jet, \figref{fig:rap},
  as well as for more tailored situations, c.f. \figref{fig:ppCF}.

In a gedanken experiment, where quarks and gluons
are collided, larger differences can be found, indicating that cancellation of subleading
$\Nc$ effects are present at the LHC. 
To capture this situation, we consider 
  LHC events while requiring that of the two most energetic jets, one is central and one is forward.
  In this case we find differences of $5-10\%$ also for standard QCD observables.
  These differences are not only in the tails of the
  distributions, but over the whole range of several observables.

Turning to soft observables the situation is different. In many cases, including
jet resolution variables at low scales,
charged particle multiplicities (\figref{fig:ee2}), individual hadron multiplicities
and the number of very soft jets at LEP, we find large effects, of several 10\%. 
While we cannot expect to make accurate predictions for any of these cases,
due to sensitivity to hadronization, multiple interaction and resummation effects,
it should be stressed that subleading $\Nc$ effects can be expected to play an important
role for the final state of the shower, entering the hadronization.
An immediate extension of this work is therefore retuning of the
$\Nc=3$ shower. Indeed, an improved description to (most) observables cannot
be expected until retuning is performed.

Another natural next step is to include virtual corrections. Virtual gluon exchanges would allow
rearrangement of the color structure without emission. This would be expected to have an effect on
gap fraction observables, such as in \figref{fig:ttbar}, and we therefore caution that our
conclusions regarding the magnitude of the subleading $\Nc$ corrections may not
be applicable in these cases.

More precisely, virtual gluon
exchanges should be the mechanism underlying (perturbative) color reconnection effects.
In the longer perspective, it would be desirable to update the standard hadronization 
model to encompass a subleading $\Nc$ shower.

For these reasons, we like to stress that this work should be considered
as the {\it start} of subleading $\Nc$ corrections at
the LHC, not the {\it end}. Indeed much work remains to be
done.

\section*{Note added}

While this work has been finalized, a similar approach has been reported in
\cite{Isaacson:2018zdi}. In \cite{Isaacson:2018zdi} color matrix element corrections,
as well as the weighted Sudakov algorithm for final state evolution is used, 
but a sampling of color structure is advocated.

\acknowledgments We thank Johannes Bellm and Mike Seymour for
discussions on color flows in dipole showers and Torbj\"orn
Sj\"ostrand for comments on the manuscript.
Johan Thor\'{e}n wishes to thank Universit\"{a}t Wien for their
hospitality.
This work was supported
by the Swedish Research Council (contract numbers 2012-02744 
and 2016-05996), 
and in
part by the MCnetITN3 H2020 Marie Curie Initial Training Network,
contract number 722104, as well as the European Union's Horizon 2020
research and innovation programme (grant agreement No 668679), and 
the COST action (``Unraveling new physics at the LHC through
the precision frontier'') No.\ CA16201.

\appendix
\section{Choices for the modified veto algorithm}
\label{app:sudakov}
This appendix contains the details for the choice of acceptance probability and overestimate kernel for the modified weighted Sudakov veto algorithm. The arguments of the splitting kernels, $p_\perp^2,z,p_{\tilde{ij}}$ and $p_{\tilde{k}}$, have been suppressed for clarity in the following equations. 

We use the modified version of the weighted sudakov veto algorithm
from \cite{Bellm:2016voq}, as described in \secref{sec:sudakov}.  The
splitting kernel we aim at generating, ${\cal P}_{ij,k}$, is given in
\eqref{eq:Pijk}. Defining ${\cal P}_{ij,k}^{\Nc\to\infty}$ to be the
leading color limit of ${\cal P}_{ij,k}$, assuming $\tilde{ij}$ and
$\tilde{k}$ are color connected (cf.\,\eqref{eq:InfCol}), and letting
${\cal R}_{ij,k}^{\Nc\to\infty}$ be the standard overestimate used in
Herwig, as obtained from the ExSample \cite{Platzer:2011dr} adapted
grid proposal, we can write
\begin{equation}
  \label{eq:Pijk2}
        {\cal P}_{ij,k}
        =
        \text{sign}(\omega_{\tilde{ij}\,\tilde{k}}^n) |\omega_{\tilde{ij}\,\tilde{k}}^n|
        (1+\delta_{\tilde{ij}})
             {\cal P}_{ij,k}^{\Nc\to\infty}
             \ .
\end{equation}
Our choice for the overestimate kernel is
${\cal R}_{ij,k}=|\omega_{\tilde{ij}\,\tilde{k}}^n|(1+\delta_{\tilde{ij}}){\cal R}_{ij,k}^{\Nc\to\infty}$ and our choice for the acceptance probability is
\begin{equation}
  \epsilon
  =
  \frac{|{\cal P}_{ij,k}|}{{\cal R}_{ij,k}}
  =
  \frac{{\cal P}_{ij,k}^{\Nc\to\infty}}{{\cal R}_{ij,k}^{\Nc\to\infty}}
  \ ,
\end{equation}
i.e. we use the same acceptance probability as for the leading $\Nc$ shower (if $\tilde{ij}$, $\tilde{k}$ were color connected). The acceptance weight is then
\begin{equation}
  \label{eq:accept}
  \text{sign}(\omega_{\tilde{ij}\,\tilde{k}}^n)\,
  =
  \left\{
  \begin{matrix}
     1 & \text{ for $\omega_{\tilde{ij}\,\tilde{k}}^n>0$} \\
    -1 & \text{ for $\omega_{\tilde{ij}\,\tilde{k}}^n<0$} \\
  \end{matrix}
  \right.
  \ ,
\end{equation}
and the veto weight is
\begin{equation}
  \label{eq:veto}
  \frac{
    {\cal R}_{ij,k}^{\Nc\to\infty} -
    \text{sign}(\omega_{\tilde{ij}\,\tilde{k}}^n) {\cal P}_{ij,k}^{\Nc\to\infty}
  }{
    {\cal R}_{ij,k}^{\Nc\to\infty} -
    {\cal P}_{ij,k}^{\Nc\to\infty}
  }
  =
  \left\{
  \begin{matrix}
     1 & \text{ for $\omega_{\tilde{ij}\,\tilde{k}}^n>0$} \\
     \frac{
       {\cal R}_{ij,k}^{\Nc\to\infty} + {\cal P}_{ij,k}^{\Nc\to\infty}
     }{
       {\cal R}_{ij,k}^{\Nc\to\infty} - {\cal P}_{ij,k}^{\Nc\to\infty}
     } & \text{ for $\omega_{\tilde{ij}\,\tilde{k}}^n<0$} \\
  \end{matrix}
  \right.
  \ .
\end{equation}
From \eqref{eq:accept} and \eqref{eq:veto} we see that trial emissions
corresponding to positive $\omega_{\tilde{ij}\,\tilde{k}}^n$ never change the
event weight.

\bibliographystyle{JHEP}  
\bibliography{ColorShower2} 

\providecommand{\href}[2]{#2}\begingroup\raggedright\begin{thebibliography}{10}

\bibitem{Sjostrand:2007gs}
T.~Sj{\"o}strand, S.~Mrenna and P.~Skands, \emph{{A Brief Introduction to
  PYTHIA 8.1}}, \href{https://doi.org/10.1016/j.cpc.2008.01.036}{\emph{Comput.
  Phys. Commun.} {\bfseries 178} (2008) 852}
  [\href{https://arxiv.org/abs/0710.3820}{{\ttfamily 0710.3820}}].

\bibitem{Bahr:2008pv}
M.~Bahr et~al., \emph{{Herwig++ Physics and Manual}},
  \href{https://doi.org/10.1140/epjc/s10052-008-0798-9}{\emph{Eur. Phys. J.}
  {\bfseries C58} (2008) 639}
  [\href{https://arxiv.org/abs/0803.0883}{{\ttfamily 0803.0883}}].

\bibitem{Bellm:2015jjp}
J.~Bellm et~al., \emph{{Herwig 7.0/Herwig++ 3.0 release note}},
  \href{https://doi.org/10.1140/epjc/s10052-016-4018-8}{\emph{Eur. Phys. J.}
  {\bfseries C76} (2016) 196}
  [\href{https://arxiv.org/abs/1512.01178}{{\ttfamily 1512.01178}}].

\bibitem{Gleisberg:2003xi}
T.~Gleisberg et~al., \emph{{SHERPA 1.alpha, a proof-of-concept version}},
  {\emph{JHEP} {\bfseries 02} (2004) 056}
  [\href{https://arxiv.org/abs/hep-ph/0311263}{{\ttfamily hep-ph/0311263}}].

\bibitem{Platzer:2011dq}
S.~Pl{\"a}tzer and M.~Sjodahl, \emph{{The Sudakov Veto Algorithm Reloaded}},
  \href{https://doi.org/10.1140/epjp/i2012-12026-x}{\emph{Eur. Phys. J. Plus}
  {\bfseries 127} (2012) 26} [\href{https://arxiv.org/abs/1108.6180}{{\ttfamily
  1108.6180}}].

\bibitem{Platzer:2011dr}
S.~Pl{\"a}tzer, \emph{{ExSample: A Library for Sampling Sudakov-Type
  Distributions}},
  \href{https://doi.org/10.1140/epjc/s10052-012-1929-x}{\emph{Eur. Phys. J.}
  {\bfseries C72} (2012) 1929}
  [\href{https://arxiv.org/abs/1108.6182}{{\ttfamily 1108.6182}}].

\bibitem{Platzer:2011bc}
S.~Pl{\"a}tzer and S.~Gieseke, \emph{{Dipole Showers and Automated NLO Matching
  in Herwig++}},  \href{https://arxiv.org/abs/1109.6256}{{\ttfamily
  1109.6256}}.

\bibitem{Platzer:2012bs}
S.~Pl{\"a}tzer, \emph{{Controlling inclusive cross sections in parton shower +
  matrix element merging}},
  \href{https://doi.org/10.1007/JHEP08(2013)114}{\emph{JHEP} {\bfseries 08}
  (2013) 114} [\href{https://arxiv.org/abs/1211.5467}{{\ttfamily 1211.5467}}].

\bibitem{Bellm:2017ktr}
J.~Bellm, S.~Gieseke and S.~Pl{\"a}tzer, \emph{{Merging NLO Multi-jet
  Calculations with Improved Unitarization}},
  \href{https://arxiv.org/abs/1705.06700}{{\ttfamily 1705.06700}}.

\bibitem{Dasgupta:2018nvj}
M.~Dasgupta, F.~A. Dreyer, K.~Hamilton, P.~F. Monni and G.~P. Salam,
  \emph{{Logarithmic accuracy of parton showers: a fixed-order study}},
  \href{https://arxiv.org/abs/1805.09327}{{\ttfamily 1805.09327}}.

\bibitem{Lonnblad:2001iq}
L.~L{\"o}nnblad, \emph{{Correcting the colour-dipole cascade model with fixed
  order matrix elements}}, {\emph{JHEP} {\bfseries 05} (2002) 046}
  [\href{https://arxiv.org/abs/hep-ph/0112284}{{\ttfamily hep-ph/0112284}}].

\bibitem{Krauss:2002up}
F.~Krauss, \emph{{Matrix elements and parton showers in hadronic
  interactions}}, {\emph{JHEP} {\bfseries 08} (2002) 015}
  [\href{https://arxiv.org/abs/hep-ph/0205283}{{\ttfamily hep-ph/0205283}}].

\bibitem{Hoche:2006ph}
S.~Hoeche, F.~Krauss, N.~Lavesson, L.~Lonnblad, M.~Mangano et~al.,
  \emph{{Matching parton showers and matrix elements}},
  \href{https://arxiv.org/abs/hep-ph/0602031}{{\ttfamily hep-ph/0602031}}.

\bibitem{Lavesson:2007uu}
N.~Lavesson and L.~L{\"o}nnblad, \emph{{Merging parton showers and matrix
  elements -- back to basics}},
  \href{https://doi.org/10.1088/1126-6708/2008/04/085}{\emph{JHEP} {\bfseries
  04} (2008) 085} [\href{https://arxiv.org/abs/0712.2966}{{\ttfamily
  0712.2966}}].

\bibitem{Hoeche:2009rj}
S.~Hoeche, F.~Krauss, S.~Schumann and F.~Siegert, \emph{{QCD matrix elements
  and truncated showers}},
  \href{https://doi.org/10.1088/1126-6708/2009/05/053}{\emph{JHEP} {\bfseries
  05} (2009) 053} [\href{https://arxiv.org/abs/0903.1219}{{\ttfamily
  0903.1219}}].

\bibitem{Hamilton:2009ne}
K.~Hamilton, P.~Richardson and J.~Tully, \emph{{A modified CKKW matrix element
  merging approach to angular-ordered parton showers}},
  \href{https://arxiv.org/abs/0905.3072}{{\ttfamily 0905.3072}}.

\bibitem{Dobbs:2001dq}
M.~Dobbs, \emph{{Phase space veto method for next-to-leading order event
  generators in hadronic collisions}},
  \href{https://doi.org/10.1103/PhysRevD.65.094011}{\emph{Phys. Rev.}
  {\bfseries D65} (2002) 094011}
  [\href{https://arxiv.org/abs/hep-ph/0111234}{{\ttfamily hep-ph/0111234}}].

\bibitem{Frixione:2002ik}
S.~Frixione and B.~R. Webber, \emph{{Matching NLO QCD computations and parton
  shower simulations}}, {\emph{JHEP} {\bfseries 06} (2002) 029}
  [\href{https://arxiv.org/abs/hep-ph/0204244}{{\ttfamily hep-ph/0204244}}].

\bibitem{Nason:2004rx}
P.~Nason, \emph{{A new method for combining NLO QCD with shower Monte Carlo
  algorithms}},
  \href{https://doi.org/10.1088/1126-6708/2004/11/040}{\emph{JHEP} {\bfseries
  11} (2004) 040} [\href{https://arxiv.org/abs/hep-ph/0409146}{{\ttfamily
  hep-ph/0409146}}].

\bibitem{Nagy:2005aa}
Z.~Nagy and D.~E. Soper, \emph{{Matching parton showers to NLO computations}},
  {\emph{JHEP} {\bfseries 10} (2005) 024}
  [\href{https://arxiv.org/abs/hep-ph/0503053}{{\ttfamily hep-ph/0503053}}].

\bibitem{Frixione:2010ra}
S.~Frixione, F.~Stoeckli, P.~Torrielli and B.~R. Webber, \emph{{NLO QCD
  corrections in Herwig++ with MC@NLO}},
  \href{https://doi.org/10.1007/JHEP01(2011)053}{\emph{JHEP} {\bfseries 1101}
  (2011) 053} [\href{https://arxiv.org/abs/1010.0568}{{\ttfamily 1010.0568}}].

\bibitem{Platzer:2012np}
S.~Pl{\"a}tzer and M.~Sjodahl, \emph{Subleading {$N_c$} improved parton
  showers}, \href{https://doi.org/10.1007/JHEP07(2012)042}{\emph{JHEP}
  {\bfseries 1207} (2012) 042}
  [\href{https://arxiv.org/abs/1201.0260}{{\ttfamily 1201.0260}}].

\bibitem{Nagy:2012bt}
Z.~Nagy and D.~E. Soper, \emph{{Parton shower evolution with subleading
  color}}, \href{https://doi.org/10.1007/JHEP06(2012)044}{\emph{JHEP}
  {\bfseries 06} (2012) 044} [\href{https://arxiv.org/abs/1202.4496}{{\ttfamily
  1202.4496}}].

\bibitem{Nagy:2015hwa}
Z.~Nagy and D.~E. Soper, \emph{{Effects of subleading color in a parton
  shower}}, \href{https://doi.org/10.1007/JHEP07(2015)119}{\emph{JHEP}
  {\bfseries 07} (2015) 119}
  [\href{https://arxiv.org/abs/1501.00778}{{\ttfamily 1501.00778}}].

\bibitem{Platzer:2013fha}
S.~Pl{\"a}tzer, \emph{{Summing Large-$N$ Towers in Colour Flow Evolution}},
  \href{https://doi.org/10.1140/epjc/s10052-014-2907-2}{\emph{Eur. Phys. J.}
  {\bfseries C74} (2014) 2907}
  [\href{https://arxiv.org/abs/1312.2448}{{\ttfamily 1312.2448}}].

\bibitem{Martinez:2018ffw}
R.~{\'A}. Mart{\'i}nez, M.~De~Angelis, J.~R. Forshaw, S.~Pl{\"a}tzer and M.~H.
  Seymour, \emph{{Soft gluon evolution and non-global logarithms}},
  \href{https://arxiv.org/abs/1802.08531}{{\ttfamily 1802.08531}}.

\bibitem{Bellm:2017bvx}
J.~Bellm et~al., \emph{{Herwig 7.1 Release Note}},
  \href{https://arxiv.org/abs/1705.06919}{{\ttfamily 1705.06919}}.

\bibitem{Platzer:2009jq}
S.~Pl{\"a}tzer and S.~Gieseke, \emph{{Coherent Parton Showers with Local
  Recoils}}, \href{https://doi.org/10.1007/JHEP01(2011)024}{\emph{JHEP}
  {\bfseries 01} (2011) 024} [\href{https://arxiv.org/abs/0909.5593}{{\ttfamily
  0909.5593}}].

\bibitem{Catani:1996jh}
S.~Catani and M.~Seymour, \emph{{The Dipole formalism for the calculation of
  QCD jet cross-sections at next-to-leading order}},
  \href{https://doi.org/10.1016/0370-2693(96)00425-X}{\emph{Phys.Lett.}
  {\bfseries B378} (1996) 287}
  [\href{https://arxiv.org/abs/hep-ph/9602277}{{\ttfamily hep-ph/9602277}}].

\bibitem{Catani:1996vz}
{S. Catani and M.H. Seymour}, \emph{{A general algorithm for calculating jet
  cross sections in NLO QCD}},
  \href{https://doi.org/10.1016/S0550-3213(96)00589-5}{\emph{Nucl. Phys.}
  {\bfseries B485} (1997) 291}
  [\href{https://arxiv.org/abs/hep-ph/9605323}{{\ttfamily hep-ph/9605323}}].

\bibitem{ttbarPaper}
K.~Cormier, S.~Pl{\"a}tzer, C.~Reuschle, P.~Richardson and S.~Webster
  {{\hspace*{-1 mm}}, in preparation}.

\bibitem{Catani:2002hc}
S.~Catani, S.~Dittmaier, M.~H. Seymour and Z.~Trocsanyi, \emph{{The Dipole
  formalism for next-to-leading order QCD calculations with massive partons}},
  \href{https://doi.org/10.1016/S0550-3213(02)00098-6}{\emph{Nucl. Phys.}
  {\bfseries B627} (2002) 189}
  [\href{https://arxiv.org/abs/hep-ph/0201036}{{\ttfamily hep-ph/0201036}}].

\bibitem{Gustafson:1987rq}
G.~Gustafson and U.~Pettersson, \emph{{Dipole Formulation of QCD Cascades}},
  \href{https://doi.org/10.1016/0550-3213(88)90441-5}{\emph{Nucl. Phys.}
  {\bfseries B306} (1988) 746}.

\bibitem{Bellm:2018wwz}
J.~Bellm, \emph{{Colour Rearrangement for Dipole Showers}}, {\emph{Eur. Phys.
  J.} {\bfseries C78} (2018) 601}
  [\href{https://arxiv.org/abs/1801.06113}{{\ttfamily 1801.06113}}].

\bibitem{Sjostrand:1985xi}
T.~Sj{\"o}strand, \emph{{A Model for Initial State Parton Showers}},
  \href{https://doi.org/10.1016/0370-2693(85)90674-4}{\emph{Phys. Lett.}
  {\bfseries 157B} (1985) 321}.

\bibitem{Paton:1969je}
J.~E. Paton and H.-M. Chan, \emph{Generalized {V}eneziano model with isospin},
  \href{https://doi.org/10.1016/0550-3213(69)90038-8}{\emph{Nucl. Phys. B}
  {\bfseries 10} (1969) 516}.

\bibitem{Berends:1987cv}
F.~A. Berends and W.~Giele, \emph{{The Six Gluon Process as an Example of
  Weyl-Van Der Waerden Spinor Calculus}},
  \href{https://doi.org/10.1016/0550-3213(87)90604-3}{\emph{Nucl. Phys.}
  {\bfseries B294} (1987) 700}.

\bibitem{Mangano:1987xk}
M.~L. Mangano, S.~J. Parke and Z.~Xu, \emph{Duality and multi-gluon
  scattering}, \href{https://doi.org/10.1016/0550-3213(88)90001-6}{\emph{Nucl.
  Phys. B} {\bfseries 298} (1988) 653}.

\bibitem{Mangano:1988kk}
M.~L. Mangano, \emph{The color structure of gluon emission},
  \href{https://doi.org/10.1016/0550-3213(88)90453-1}{\emph{Nucl. Phys. B}
  {\bfseries 309} (1988) 461}.

\bibitem{Kosower:1988kh}
D.~A. Kosower, \emph{{Color Factorization for Fermionic Amplitudes}},
  \href{https://doi.org/10.1016/0550-3213(89)90361-1}{\emph{Nucl. Phys.}
  {\bfseries B315} (1989) 391}.

\bibitem{Nagy:2007ty}
Z.~Nagy and D.~E. Soper, \emph{Parton showers with quantum interference},
  \href{https://doi.org/10.1088/1126-6708/2007/09/114}{\emph{JHEP} {\bfseries
  09} (2007) 114} [\href{https://arxiv.org/abs/0706.0017}{{\ttfamily
  0706.0017}}].

\bibitem{Sjodahl:2009wx}
M.~Sjodahl, \emph{Color structure for soft gluon resummation -- a general
  recipe}, \href{https://doi.org/10.1088/1126-6708/2009/09/087}{\emph{JHEP}
  {\bfseries 0909} (2009) 087}
  [\href{https://arxiv.org/abs/0906.1121}{{\ttfamily 0906.1121}}].

\bibitem{Alwall:2011uj}
J.~Alwall, M.~Herquet, F.~Maltoni, O.~Mattelaer and T.~Stelzer, \emph{{MadGraph
  5 : Going Beyond}},
  \href{https://doi.org/10.1007/JHEP06(2011)128}{\emph{JHEP} {\bfseries 1106}
  (2011) 128} [\href{https://arxiv.org/abs/1106.0522}{{\ttfamily 1106.0522}}].

\bibitem{Sjodahl:2014opa}
M.~Sjodahl, \emph{{ColorFull -- a C++ library for calculations in SU(Nc) color
  space}}, \href{https://doi.org/10.1140/epjc/s10052-015-3417-6}{\emph{Eur.
  Phys. J.} {\bfseries C75} (2015) 236}
  [\href{https://arxiv.org/abs/1412.3967}{{\ttfamily 1412.3967}}].

\bibitem{Kyrieleis:2005dt}
A.~Kyrieleis and M.~H. Seymour, \emph{The colour evolution of the process $qq
  \to qqg$}, {\emph{JHEP} {\bfseries 01} (2006) 085}
  [\href{https://arxiv.org/abs/hep-ph/0510089}{{\ttfamily hep-ph/0510089}}].

\bibitem{Dokshitzer:2005ig}
Y.~L. Dokshitzer and G.~Marchesini, \emph{Soft gluons at large angles in hadron
  collisions}, {\emph{JHEP} {\bfseries 01} (2006) 007}
  [\href{https://arxiv.org/abs/hep-ph/0509078}{{\ttfamily hep-ph/0509078}}].

\bibitem{Sjodahl:2008fz}
M.~Sjodahl, \emph{Color evolution of 2 $\to$ 3 processes},
  \href{https://doi.org/10.1088/1126-6708/2008/12/083}{\emph{JHEP} {\bfseries
  12} (2008) 083} [\href{https://arxiv.org/abs/0807.0555}{{\ttfamily
  0807.0555}}].

\bibitem{Beneke:2009rj}
M.~Beneke, P.~Falgari and C.~Schwinn, \emph{Soft radiation in heavy-particle
  pair production: {A}ll-order colour structure and two-loop anomalous
  dimension},
  \href{https://doi.org/10.1016/j.nuclphysb.2009.11.004}{\emph{Nucl. Phys. B}
  {\bfseries 828} (2010) 69} [\href{https://arxiv.org/abs/0907.1443}{{\ttfamily
  0907.1443}}].

\bibitem{Keppeler:2012ih}
S.~Keppeler and M.~Sjodahl, \emph{{Orthogonal multiplet bases in SU(Nc) color
  space}}, \href{https://doi.org/10.1007/JHEP09(2012)124}{\emph{JHEP}
  {\bfseries 09} (2012) 124} [\href{https://arxiv.org/abs/1207.0609}{{\ttfamily
  1207.0609}}].

\bibitem{Du:2015apa}
Y.-J. Du, M.~Sjodahl and J.~Thor{\'e}n, \emph{{Recursion in multiplet bases for
  tree-level MHV gluon amplitudes}},
  \href{https://doi.org/10.1007/JHEP05(2015)119}{\emph{JHEP} {\bfseries 05}
  (2015) 119} [\href{https://arxiv.org/abs/1503.00530}{{\ttfamily
  1503.00530}}].

\bibitem{Sjodahl:2015qoa}
M.~Sjodahl and J.~Thor{\'e}n, \emph{{Decomposing color structure into multiplet
  bases}}, \href{https://doi.org/10.1007/JHEP09(2015)055}{\emph{JHEP}
  {\bfseries 09} (2015) 055}
  [\href{https://arxiv.org/abs/1507.03814}{{\ttfamily 1507.03814}}].

\bibitem{'tHooft:1973jz}
G.~'t~Hooft, \emph{{A PLANAR DIAGRAM THEORY FOR STRONG INTERACTIONS}},
  \href{https://doi.org/10.1016/0550-3213(74)90154-0}{\emph{Nucl. Phys.}
  {\bfseries B72} (1974) 461}.

\bibitem{Kanaki:2000ms}
A.~Kanaki and C.~G. Papadopoulos, \emph{{HELAC-PHEGAS: Automatic computation of
  helicity amplitudes and cross-sections}},
  \href{https://arxiv.org/abs/hep-ph/0012004}{{\ttfamily hep-ph/0012004}}.

\bibitem{Maltoni:2002mq}
F.~Maltoni, K.~Paul, T.~Stelzer and S.~Willenbrock, \emph{Color flow
  decomposition of {QCD} amplitudes},
  \href{https://doi.org/10.1103/PhysRevD.67.014026}{\emph{Phys. Rev. D}
  {\bfseries 67} (2003) 014026}
  [\href{https://arxiv.org/abs/hep-ph/0209271}{{\ttfamily hep-ph/0209271}}].

\bibitem{Bellm:2016voq}
J.~Bellm, S.~Pl{\"a}tzer, P.~Richardson, A.~Si{\'o}dmok and S.~Webster,
  \emph{{Reweighting Parton Showers}},
  \href{https://doi.org/10.1103/PhysRevD.94.034028}{\emph{Phys. Rev.}
  {\bfseries D94} (2016) 034028}
  [\href{https://arxiv.org/abs/1605.08256}{{\ttfamily 1605.08256}}].

\bibitem{Cacciari:2008gp}
M.~Cacciari, G.~P. Salam and G.~Soyez, \emph{{The Anti-k(t) jet clustering
  algorithm}}, \href{https://doi.org/10.1088/1126-6708/2008/04/063}{\emph{JHEP}
  {\bfseries 04} (2008) 063} [\href{https://arxiv.org/abs/0802.1189}{{\ttfamily
  0802.1189}}].

\bibitem{Buckley:2010ar}
A.~Buckley, J.~Butterworth, L.~Lonnblad, D.~Grellscheid, H.~Hoeth, J.~Monk
  et~al., \emph{{Rivet user manual}},
  \href{https://doi.org/10.1016/j.cpc.2013.05.021}{\emph{Comput. Phys. Commun.}
  {\bfseries 184} (2013) 2803}
  [\href{https://arxiv.org/abs/1003.0694}{{\ttfamily 1003.0694}}].

\bibitem{Isaacson:2018zdi}
J.~Isaacson and S.~Prestel, \emph{{On stochastically sampling color
  configurations}},  \href{https://arxiv.org/abs/1806.10102}{{\ttfamily
  1806.10102}}.

\bibitem{Abreu:1996na}
{\scshape DELPHI} collaboration, P.~Abreu et~al., \emph{{Tuning and test of
  fragmentation models based on identified particles and precision event shape
  data}}, \href{https://doi.org/10.1007/s002880050295}{\emph{Z. Phys.}
  {\bfseries C73} (1996) 11}.

\bibitem{Barate:1996fi}
{\scshape ALEPH} collaboration, R.~Barate et~al., \emph{{Studies of quantum
  chromodynamics with the ALEPH detector}},
  \href{https://doi.org/10.1016/S0370-1573(97)00045-8}{\emph{Phys. Rept.}
  {\bfseries 294} (1998) 1}.

\bibitem{Heister:2003aj}
{\scshape ALEPH} collaboration, A.~Heister et~al., \emph{{Studies of QCD at e+
  e- centre-of-mass energies between 91-GeV and 209-GeV}},
  \href{https://doi.org/10.1140/epjc/s2004-01891-4}{\emph{Eur. Phys. J.}
  {\bfseries C35} (2004) 457}.

\bibitem{radek.simon}
S.~Gieseke, S.~Pl{\"a}tzer and R.~Podskubka in preparation.

\bibitem{Khachatryan:2011dx}
{\scshape CMS} collaboration, V.~Khachatryan et~al., \emph{{First Measurement
  of Hadronic Event Shapes in $pp$ Collisions at $\sqrt {s}=7$ TeV}},
  \href{https://doi.org/10.1016/j.physletb.2011.03.060}{\emph{Phys. Lett.}
  {\bfseries B699} (2011) 48}
  [\href{https://arxiv.org/abs/1102.0068}{{\ttfamily 1102.0068}}].

\bibitem{Aad:2012meb}
{\scshape ATLAS} collaboration, G.~Aad et~al., \emph{{ATLAS Measurements of the
  Properties of Jets for Boosted Particle Searches}},
  \href{https://doi.org/10.1103/PhysRevD.86.072006}{\emph{Phys. Rev.}
  {\bfseries D86} (2012) 072006}
  [\href{https://arxiv.org/abs/1206.5369}{{\ttfamily 1206.5369}}].

\bibitem{ATLAS:2012al}
{\scshape ATLAS} collaboration, G.~Aad et~al., \emph{{Measurement of $t
  \bar{t}$ production with a veto on additional central jet activity in pp
  collisions at sqrt(s) = 7 TeV using the ATLAS detector}},
  \href{https://doi.org/10.1140/epjc/s10052-012-2043-9}{\emph{Eur. Phys. J.}
  {\bfseries C72} (2012) 2043}
  [\href{https://arxiv.org/abs/1203.5015}{{\ttfamily 1203.5015}}].

\bibitem{Aad:2013fba}
{\scshape ATLAS} collaboration, G.~Aad et~al., \emph{{Measurement of jet shapes
  in top-quark pair events at $\sqrt{s}$ = 7 TeV using the ATLAS detector}},
  \href{https://doi.org/10.1140/epjc/s10052-013-2676-3}{\emph{Eur. Phys. J.}
  {\bfseries C73} (2013) 2676}
  [\href{https://arxiv.org/abs/1307.5749}{{\ttfamily 1307.5749}}].

\bibitem{Chatrchyan:2013fha}
{\scshape CMS} collaboration, S.~Chatrchyan et~al., \emph{{Probing color
  coherence effects in pp collisions at $\sqrt{s}=7\,\text {TeV} $}},
  \href{https://doi.org/10.1140/epjc/s10052-014-2901-8}{\emph{Eur. Phys. J.}
  {\bfseries C74} (2014) 2901}
  [\href{https://arxiv.org/abs/1311.5815}{{\ttfamily 1311.5815}}].

\bibitem{Aad:2014pua}
{\scshape ATLAS} collaboration, G.~Aad et~al., \emph{{Measurements of jet
  vetoes and azimuthal decorrelations in dijet events produced in $pp$
  collisions at $\sqrt{s}=7\,\mathrm{TeV}$ using the ATLAS detector}},
  \href{https://doi.org/10.1140/epjc/s10052-014-3117-7}{\emph{Eur. Phys. J.}
  {\bfseries C74} (2014) 3117}
  [\href{https://arxiv.org/abs/1407.5756}{{\ttfamily 1407.5756}}].

\bibitem{Aad:2016ria}
{\scshape ATLAS} collaboration, G.~Aad et~al., \emph{{Measurement of
  event-shape observables in $Z \rightarrow \ell ^{+} \ell ^{-}$ events in $pp$
  collisions at $\sqrt{s}=$ 7 TeV with the ATLAS detector at the LHC}},
  \href{https://doi.org/10.1140/epjc/s10052-016-4176-8}{\emph{Eur. Phys. J.}
  {\bfseries C76} (2016) 375}
  [\href{https://arxiv.org/abs/1602.08980}{{\ttfamily 1602.08980}}].

\bibitem{Banfi:2010xy}
A.~Banfi, G.~P. Salam and G.~Zanderighi, \emph{{Phenomenology of event shapes
  at hadron colliders}},
  \href{https://doi.org/10.1007/JHEP06(2010)038}{\emph{JHEP} {\bfseries 06}
  (2010) 038} [\href{https://arxiv.org/abs/1001.4082}{{\ttfamily 1001.4082}}].

\bibitem{Forshaw:2007vb}
J.~R. Forshaw and M.~Sjodahl, \emph{{Soft gluons in Higgs plus two jet
  production}},
  \href{https://doi.org/10.1088/1126-6708/2007/09/119}{\emph{JHEP} {\bfseries
  09} (2007) 119} [\href{https://arxiv.org/abs/0705.1504}{{\ttfamily
  0705.1504}}].

\bibitem{Cox:2010ug}
B.~E. Cox, J.~R. Forshaw and A.~D. Pilkington, \emph{{Extracting Higgs boson
  couplings using a jet veto}},
  \href{https://doi.org/10.1016/j.physletb.2010.12.011}{\emph{Phys. Lett.}
  {\bfseries B696} (2011) 87}
  [\href{https://arxiv.org/abs/1006.0986}{{\ttfamily 1006.0986}}].

\bibitem{DuranDelgado:2011tp}
R.~M. Duran~Delgado, J.~R. Forshaw, S.~Marzani and M.~H. Seymour, \emph{{The
  dijet cross section with a jet veto}},
  \href{https://doi.org/10.1007/JHEP08(2011)157}{\emph{JHEP} {\bfseries 08}
  (2011) 157} [\href{https://arxiv.org/abs/1107.2084}{{\ttfamily 1107.2084}}].

\end{thebibliography}\endgroup

\end{document}